\documentclass[twocolumn,american]{article}
\usepackage[T1]{fontenc}
\UseRawInputEncoding
\usepackage{geometry}
\usepackage{xcolor}
\usepackage{units}
\usepackage{amsmath}
\usepackage{amsthm}
\usepackage{amssymb}
\usepackage{graphicx}
\usepackage{wasysym}
\PassOptionsToPackage{normalem}{ulem}
\usepackage{ulem}

\makeatletter

\providecolor{lyxadded}{rgb}{0,0,1}
\providecolor{lyxdeleted}{rgb}{1,0,0}
\DeclareRobustCommand{\mklyxadded}[1]{\bgroup\color{lyxadded}{}#1\egroup}
\DeclareRobustCommand{\mklyxdeleted}[1]{\bgroup\color{lyxdeleted}\mklyxsout{#1}\egroup}
\DeclareRobustCommand{\mklyxsout}[1]{\ifx\\#1\else\sout{#1}\fi}

\usepackage{abstract} 
\newcommand{\makeabstract}{\@ifundefined{abstractcontent}{}{\begin{abstract}\abstractcontent\end{abstract}}}
\newcommand{\makefrontmatter}{\if@twocolumn{\twocolumn[\maketitle\makeabstract\vskip2\baselineskip]\saythanks}\else{\maketitle\makeabstract}\fi}

\usepackage{amsmath}
\usepackage[bb=dsserif]{mathalpha}
\usepackage{bm}
\usepackage[latin1]{inputenc}
\usepackage{csquotes}

\makeatother

\usepackage{babel}
\usepackage[bibstyle=numeric,citestyle=numeric-comp-archaeology,sorting=none]{biblatex}
\begin{document}
\title{Quantifying imperfect cognition via achieved information gain}
\author{Torsten En{\ss}lin \\
{\small$^{1}$ Max Planck Institute for Astrophysics, Karl-Schwarzschild-Str.\ 1,
85748 Garching, Germany}\\
{\small$^{2}$ Deutsches Zentrum für Astrophysik, Postplatz 1, 02826
Görlitz, Germany}\\
{\small$^{3}$ Ludwig-Maximilians-Universit\"at M\"unchen, Geschwister-Scholl-Platz
1, 80539 Munich, Germany}\\
{\small$^{4}$ Excellence Cluster ORIGINS, Boltzmannstr.\ 2, 85748
Garching, Germany}}
\newcommand{\abstractcontent}{Cognition, information processing in form of inference, communication,
and memorization, is the central activity of any intelligence. Its
physical realization in a brain, computer, or in any other intelligent
system requires resources like time, energy, memory, bandwidth, money,
and others. Due to limited resources, many real world intelligent
systems perform only imperfect cognition. To understand the trade-off
between accuracy and resource investments in existing systems, e.g.\ in
biology, as well as for the resource-aware optimal design of information
processing systems, like computer algorithms and artificial neural
networks, a quantification of information obtained in an imperfect
cognitive operation is desirable. To this end, we propose the concept
of the \emph{achieved information gain} (AIG) of a belief update,
which is given by the amount of information obtained by updating from
the initial state of knowledge to the ideal state, minus the amount
that a change from the imperfect to the ideal state would yield. AIG
has many desirable properties for quantifying imperfect cognition.
The ratio of achieved to ideally obtainable information measures \emph{cognitive
fidelity} and that of AIG to the necessary cognitive effort measures
\emph{cognitive efficiency}. This work provides an axiomatic derivation
of AIG, relates it to other information measures, illustrates its
application to common scenarios of posterior inaccuracies, and discusses
the implication of cognitive efficiency for sustainable resource allocation
in computational inference.\linebreak{}
\textbf{\emph{Key words:}} \emph{information theory; communication
theory; entropy; sustainable computing }}

\makefrontmatter
\vspace{0cm}

\section{Introduction}

\subsection{Information measures}

Any information processing entity in the physical world -- cognitive
system for short -- has to operate with limited resources. This is
valid for technical as well as for biological systems, which therefore
all need to make trade-offs between accuracy and costs. In order to
assess the former, measures of the amount of information gained, lost,
or transmitted by any information processing operation are needed.
Fundamental cognitive operations we consider here embrace information
transmission, memorization, and inference.

The commonly used measure for the amount of information are entropy
and in particular relative entropy. Relative entropy, aka Kullback-Leibler
divergence \cite{10.1214/aoms/1177729694}, roots in statistical mechanics
\parencite{gibbs1906thermodynamics} and information theory \cite{shannon}
and can be derived in a number of ways \cites{mccarthy1956measures}{Bernardo}{winkler1968good}{caticha2006updating}{gneiting2007strictly}{knuth2012foundations}{2017Entrp..19..402L}{Harremoes}{gkelsinis2020theoretical}{2024AnP...53600334E}.
It characterizes the amount of information gained by changing from
a less informed initial state of knowledge to a more informed one,
or the amount of information lost in the reverse change. It is therefore
central to the understanding, characterization, and design of information
processing systems and consequently can be found in a large number
of contexts. To name a few: The Maximum Entropy Principle to assign
probability distributions \cite{PhysRev.106.620,PhysRev.108.171,jaynes1963information,jaynes1968prior,jaynes03},
approximate Bayesian reasoning via Expectation Propagation \parencite{minka2013expectation}
and Variational Inference \parencite{blei2017variational,knollmuller2019metric,2021Entrp..23..853F},
information geometry \parencite{amari1997information,ay2017information},
mean field approximations in inference \parencites{opper2001advanced}{10.7551/mitpress/1100.003.0007}{10.7551/mitpress/1100.003.0014}{10.7551/mitpress/1100.003.0020}{10.7551/mitpress/1100.003.0021},
training of neural networks \parencite{graves2011practical}, in particular
of Variational Autoencoders \parencite{kingma2014stochastic,kingma2019introduction,milosevic2021bayesian,zacherl2021probabilistic},
optimal coding \parencite{shannon1959probability}, lossy data compression
\parencites{salomon2002data}{harth2021toward}, model fusion \parencite{claici2020model},
causal inference based on mutual information \parencites{JMLR:v17:15-420}{nogueira2022methods},
information field theory \parencite{2010PhRvE..82e1112E}, active
learning and active inference \parencites{buckley2017free}{da2020active}{10.1162/neco_a_01351}{2024arXiv240214460C},
ecology \cite{burnham2001kullback}, and computational psychology
\parencite{sun2008cambridge,2022AnP...53400277E}. As diverse as this
list is, it is far from being complete.

\begin{figure}
\includegraphics[clip,width=1\columnwidth]{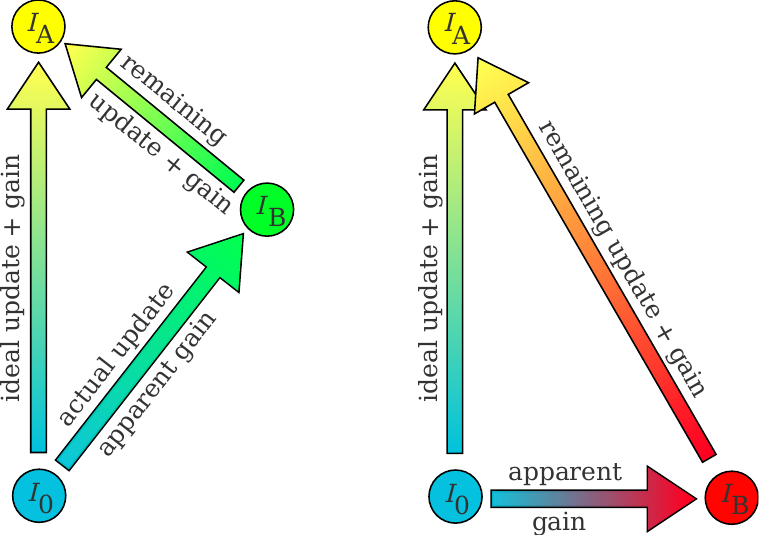}\caption{AIG as the information gain of the ideal update $I_{\text{0}}\rightarrow I_{\text{A}}$
minus that of the remaining one $I_{\text{B}}\rightarrow I_{\text{A}}$.
The lengths of the arrows indicate the amount of information that
seemed to be gained between target and start states. Scenario with
positive (left) and negative (right) AIG are shown. Note that the
apparent information gain from the actual update $I_{\text{0}}\rightarrow I_{\text{B}}$
is irrelevant for the AIG. \protect\label{fig:Realized-information-gain}}
\end{figure}

The problem of relative entropy as a measure of the amount of information
obtained in a cognitive update is that it is insensitive to whether
the update went into the right or the wrong direction. Becoming very
sure about something that is wrong comes with a significant positive
relative entropy, despite it should rather be associated with negative
information, as undoing the wrong update will require a positive amount
of information just to restore the initial state, and that had no
information gain. The relative entropy between updated and initial
knowledge state therefore only characterizes the apparent information
gain, not the real one. The real one should also take into account
how much of the update goes into the right direction, in order to
be able to discriminate purely apparent from actual information. It
therefore depends on three information states, the initial, the final,
and the ideal one, see Fig.\ \ref{fig:Realized-information-gain}.
We argue in this work that using the relative entropy of the ideal
update minus that of the remaining update to the ideal information
state as a measure of \emph{achieved information gain }(AIG) is a
very good way to quantify the information gained in an imperfect cognitive
operation. The AIG is zero, if there was no update, it is maximal,
if the update is ideal, and it becomes negative when the update goes
into the wrong direction. Its units are nits or bits and it has a
simple intuitive interpretation: It provides an estimate of the reduction
in surprise due to the actual update, calculated from the perspective
of the ideal knowledge state.

AIG measures the amount of information gained by approximate cognitive
operations. It therefore allows to characterize the \emph{cognitive
fidelity} (CF) of such an operation as the ratio of the achieved to
ideal information gain and its \emph{cognitive efficiency} (CE) as
the ratio of AIG to invested resources like time, money, energy, and
environmental footprint.

The here proposed definition of CE seems also to be well aligned with
that used in psychology and cognitive research: ``Cognitive efficiency
(CE) is generally defined as qualitative increases in knowledge gained
in relation to the time and effort invested in knowledge acquisition``
\cite{HOFFMAN2012133}. The improvement of the quality of knowledge
indicates that only knowledge in the right direction should be accounted
for CE, exactly as done in our definitions of AIG and CE.

AIG, CE, and CF should have important technological applications.
They can guide the decision on which method to choose out of a number
of data processing methodologies that differ in terms of fidelity
and computational costs. As computational costs of data analysis can
be substantial \parencite{2024EPJST.tmp..399B}, less expensive methods
might look advantageous at first sight. However, these might require
larger data sets in order to provide results comparable to that of
more expensive, but higher fidelity methods. As the latter imply less
measurement costs, they might actually be cheaper from a global perspective,
despite their larger computational costs. In order to judge, the information
gain as a function of the data set size needs to be quantified for
every method under consideration. The concepts of AIG make this possible,
and those of CF and CE derived from it can provide valuable quantitative
guidance for decisions on sustainable computing.

\subsection{Structure of the work}

This work is structured as follows: Sect.\ \ref{sec:Information-gain}
states the mathematical preliminaries, develops AIG at the cognitive
operations of communication, inference, and memorization, and defines
CF mathematically. Sect.\ \ref{sec:Axiomatic-derivation} provides
an axiomatic derivation of AIG and shows that it becomes a separable
quantity in case that initial and updated knowledge states were both
separable. Sect.\ \ref{sec:Relation-to-other} discusses the relation
of AIG to other information measures. Sect.\ \ref{sec:Instructive-examples}
illustrates the usage of AIG at a number of illustrative cases, like
updates of Bernoulli, binomial, Poisson, and Gaussian distributions
with inaccurate parameters, the neglection of cross correlations between
parameters in mean field approximations, the incomplete usage of data,
and shows how AIG can be estimated for non-Gaussian probability distributions.
Sect.\ \ref{sec:Sustainable-data-analysis} introduces the mathematical
definition of CE based on AIG and shows how CE can guide sustainable
data analysis. And finally, Sect.\ \ref{sec:Conclusions} concludes
this work with a brief summary and outlook.

\section{Information gain\protect\label{sec:Information-gain}}

\subsection{Mathematical preliminaries}

All cognitive operations modify the internal information state $I\in\mathcal{I}\equiv\mathbb{R}^{n}$
of a cognitive system, for example its memory. $\mathcal{I}$ is the
space of all possible information states, to which a generalized distance
measure will be introduced that will serve as quantifying the amount
of information gained in an update between states. A given knowledge
state $I\in\mathcal{I}$ consists of various variables (coordinates)
that encode knowledge about some situation (or signal) $s\in\mathcal{S}$
out of a measurable set $\mathcal{S}$ of possible situations. We
assume that knowing which of those situations is the case is of relevance
to the cognitive system. As it usually has only imperfect information
-- $I$ does not determine $s$ fully -- the remaining uncertainty
has to be characterized with the help of a probability distribution
over $\mathcal{S}$. Let $P(\cdot|\mathcal{S},I,M):\mathbb{P}(\mathcal{S})\mapsto[0,1]$
be the probability of a continuous quantity $s$ to be in some subset
$\mathcal{S}'\in\mathbb{P}(\mathcal{S})$ of $\mathcal{S}$. Here,
$\mathbb{P}(\mathcal{S})$ denotes the superset of $\mathcal{S}$,
the set of all of its subsets. Thus, $s\in\mathcal{S}'\subset\mathcal{S}$.
$I$ denotes the information given on $s$ and $M$ the generic (world)
model, which specifies the relation of $s$ and $I$. This probability
can be expressed as
\begin{equation}
P(\mathcal{S}'|\mathcal{S},I,M)=\int_{\mathcal{S}'}\text{d}s\,\mathcal{P}(s|\mathcal{S},I,M),
\end{equation}
with $\mathcal{P}(s|\mathcal{S},I,M)\ge0$ being a probability density
function. The latter has to obey normalization,
\begin{equation}
\int_{\mathcal{S}}\text{d}s\,\mathcal{P}(s|\mathcal{S},I,M)=1.
\end{equation}

Discrete sets of possibilities can be treated analogously, by the
replacement of the integrals by corresponding sums over $\mathcal{S}.$
As we assume in the following only one set $\mathcal{S}$ and one
model $M$ are present, we suppress these in the notation of probabilities
and define $\mathcal{P}(s|I):=\mathcal{P}(s|\mathcal{S},I,M)$.

Denoting with $I_{\text{A}}$ and $I_{\text{B}}$ two different information
states, one defines their relative entropy \parencites{10.1214/aoms/1177729694}{Bernardo}
as 
\begin{eqnarray}
\mathcal{D_{\mathcal{S}}}(I_{\text{A}},I_{\text{B}}) & := & \int_{\mathcal{S}}\text{d}s\,\mathcal{P}(s|I_{\text{A}})\ln\frac{\mathcal{P}(s|I_{\text{A}})}{\mathcal{P}(s|I_{\text{B}})}.\label{eq:RIG}
\end{eqnarray}
``$\text{A}$'' stands here for Alice and ``$\text{B}$'' for
Bob, the canonical names for the involved agents in communication
theory.  Relative entropy can be regarded as the amount of information
lost when changing from $I_{\text{A}}$ to $I_{\text{B}}$, or as
the amount gained when changing from $I_{\text{B}}$ to $I_{\text{A}}$.
It is a positive quantity, 
\begin{equation}
\mathcal{D_{\mathcal{S}}}(I_{\text{A}},I_{\text{B}})\ge0,\label{eq:KL_is_positive}
\end{equation}
with equality if and only if $I_{\text{A}}=I_{\text{B}}$.

The units of relative entropy as defined above are nits. They are
bits if in its definition instead of $\ln$, the natural logarithm,
$\log_{2}$, the logarithm to the basis two, is used. The conversion
between these units is $\text{bit}=\ln2\text{ nit}\approx0.69\text{ nit}.$

For example, learning the state $s\in\mathcal{S}=\{\text{head, tail}\}$
of a coin provides exactly one bit of information. To be specific,
let us assume that Alice knows that $s=\text{head}$, which means
$P(\text{head}|I_{\text{A}})=1$ and $P(\text{tail}|I_{\text{A}})=0$.
Bob be initially uninformed, so that his knowledge state $I_{\text{B}}$
is characterized by $P(\text{head}|I_{\text{B}})=P(\text{tail}|I_{\text{B}})=\nicefrac{1}{2}$.
This leads to an information gain by changing from $I_{\text{B}}$
to $I_{\text{A}}$ of one bit:

\begin{eqnarray}
\frac{\mathcal{D_{\mathcal{S}}}(I_{\text{A}},I_{\text{B}})}{\text{bit}} & = & \sum_{s\in\mathcal{S}}\,P(s|I_{\text{A}})\log_{2}\frac{P(s|I_{\text{A}})}{P(s|I_{\text{B}})}\\
 & = & P(\text{head}|I_{\text{A}})\log_{2}\frac{P(\text{head}|I_{\text{A}})}{P(\text{head}|I_{\text{B}})}+\nonumber \\
 &  & P(\text{\text{tail}}|I_{\text{A}})\log_{2}\frac{P(\text{\text{tail}}|I_{\text{A}})}{P(\text{tail}|I_{\text{B}})}\nonumber \\
 & = & 1\times\log_{2}\frac{1}{\nicefrac{1}{2}}+0\,\times\log_{2}\frac{0}{\nicefrac{1}{2}}=\log_{2}2=1.\nonumber 
\end{eqnarray}
Here, the identity $0\log_{2}0\equiv\lim_{\varepsilon\rightarrow0}\varepsilon\log_{2}\varepsilon=0$
is used.

Relative entropy is an asymmetric distance measure, a \emph{divergence},
with $\mathcal{D_{\mathcal{S}}}(I_{\text{A}},I_{\text{B}})\neq\mathcal{D_{\mathcal{S}}}(I_{\text{B}},I_{\text{A}})$
in general. When used to describe the loss of information, the initial
information state has to be the first argument. When quantifying the
gain of information, the final information state is the first argument.
In general, the presumably better informed state is the first argument
since this is the one over which averages are calculated. This asymmetry
roots in relative entropy serving as a scoring function that guides
the communication of Alice, whoses perspective is the basis of the
involved expectation estimate \parencite{Bernardo1979ExpectedIA,2017Entrp..19..402L,2024AnP...53600334E}.

Now we can discuss the cognitive operations of information transmission,
memorization, and inference from an information theoretical perspective.

\subsection{Information transmission}

For information transmission, or short \emph{communication}, different
amounts of information can be defined between the three involved information
states, that of the sender Alice, $I_{\text{A}}$, and the two of
the receiver Bob before and after the communication, $I_{\text{0}}$
and $I_{\text{B}}$, respectively. Here, we only consider the entropies
that have the presumably better knowledge state as their first argument,
and this will be nearly exclusively that of Alice, $I_{\text{A}}$.
$I_{\text{0}}$ always denotes the initial knowledge state of Bob,
and $I_{\text{B}}$ his state after his update due to the communication
he received from Alice. See Fig.\ \ref{fig:Realized-information-gain}
for an illustration of these three states and the three relevant relative
entropies.

The relative entropy $\mathcal{D_{\mathcal{S}}}(I_{\text{A}},I_{\text{B}})$
between the sender's belief state $I_{\text{A}}$ and the receiver's
updated belief state $I_{\text{B}}$ quantifies the amount of information
lost in the communication from A's perspective. An optimal information
transmission system tries to minimize this information loss from Alice
to Bob. This quantity can be regarded as the \textbf{remaining information
gain}, as Bob could still gain this amount of information by receiving
further messages from Alice.

Another amount of information can be defined between Bob's initial
state, denoted by $I_{0}$, and its updated version $I_{\text{B}}$.
The relative entropy of receiver's final and initial beliefs, $\mathcal{D_{\mathcal{S}}}(I_{\text{B}},I_{0})$,
quantifies the amount of information the receiver seems to have gained,
the \textbf{apparent information gain}. Since the communication does
not necessarily informs the receiver perfectly, as $\mathcal{D_{\mathcal{S}}}(I_{\text{A}},I_{\text{B}})>0$
could be the case, this information gain $\mathcal{D_{\mathcal{S}}}(I_{\text{B}},I_{0})$
is -- however -- only apparent in the eyes of Bob, but not necessarily
an actual gain of correct information from the perspective of Alice.

For example, in the coin example, Alice might have accidentally or
intentionally made Bob belief that the state of the coin is $\text{tail}$,
despite knowing it to be $\text{head}$. In this case, the apparent
information gain $\mathcal{D_{\mathcal{S}}}(I_{\text{B}},I_{0})$
is still one bit, although Bob's knowledge got worse.

The third relevant measure is the \textbf{ideal information gain}
$\mathcal{D_{\mathcal{S}}}(I_{\text{A}},I_{\text{0}})$ Bob could
have obtained in a perfect communication. The ideal information gain
serves as a benchmark for comparison with the remaining information
gain.

In order to characterize the information gain of Bob in a way that
does not have the flaw of the apparent information gain to be positive
under misinformation, we introduce the \textbf{achieved information
gain} (AIG), $\mathcal{D_{\mathcal{S}}}(I_{\text{A}},I_{\text{B}},I_{\text{0}})$.
This shall be Bob's gain after a perfect communication, $\mathcal{D_{\mathcal{S}}}(I_{\text{A}},I_{0})$,
minus the remaining amount of information to be gained, $\mathcal{D_{\mathcal{S}}}(I_{\text{A}},I_{\text{B}})$.
Bob's apparent information gain $\mathcal{D_{\mathcal{S}}}(I_{\text{B}},I_{\text{0}})$
does not play any direct role in it, as it is not a relevant measure
from Alice's perspective. The achieved information,
\begin{eqnarray}
\mathcal{D_{\mathcal{S}}}(I_{\text{A}},I_{\text{B}},I_{\text{0}}) & := & \mathcal{D_{\mathcal{S}}}(I_{\text{A}},I_{\text{0}})-\mathcal{D_{\mathcal{S}}}(I_{\text{A}},I_{\text{B}})\nonumber \\
 & = & \int_{S}\text{d}s\,\mathcal{P}(s|I_{\text{A}})\ln\frac{\mathcal{P}(s|I_{\text{B}})}{\mathcal{P}(s|I_{\text{0}})},
\end{eqnarray}
is a relative entropy-like expression that involves all three belief
states relevant in the communication, that of Alice, the sender of
the message, and those of Bob, the message's receiver, before and
after the message transmission, $I_{\text{A}}$, $I_{\text{0}}$,
and $I_{B}$, respectively. We will derive it axiomatically in Sect.\ \ref{sec:Axiomatic-derivation}.
But first, we motivate it by showing that it has a number of desirable
properties.

As relative entropy is a positive quantity, and Alice can only affect
$I_{\text{B}}$, the maximal relative information gain is given when
$\mathcal{D_{\mathcal{S}}}(I_{\text{A}},I_{\text{B}})=0$, which implies
$I_{\text{B}}=I_{\text{A}}$. In this case Bob's update is perfect,
$\mathcal{D_{\mathcal{S}}}(I_{\text{A}},I_{\text{A}},I_{\text{0}})=\mathcal{D_{\mathcal{S}}}(I_{\text{A}},I_{\text{0}})-\mathcal{D_{\mathcal{S}}}(I_{\text{A}},I_{\text{A}})=\mathcal{D_{\mathcal{S}}}(I_{\text{A}},I_{\text{0}})$,
and thus the AIG equals the ideal one. In case there is no update,
$I_{\text{B}}=I_{0}$, the AIG vanishes, as $\mathcal{D_{\mathcal{S}}}(I_{\text{A}},I_{\text{0}},I_{\text{0}})=\mathcal{D_{\mathcal{S}}}(I_{\text{A}},I_{\text{0}})-\mathcal{D_{\mathcal{S}}}(I_{\text{A}},I_{\text{0}})=0$.
Note, there can be a negative information gain in case $\mathcal{D_{\mathcal{S}}}(I_{\text{A}},I_{\text{0}})<\mathcal{D_{\mathcal{S}}}(I_{\text{A}},I_{\text{B}})$,
which happens if the updated state diverges more from the ideal than
the initial one. This would be a clear case of cognition going the
wrong direction. A situation with a positive as well as one with a
negative AIG is depicted in Fig.\ \ref{fig:Realized-information-gain}.

In case Alice informs Bob correctly about the state of a coin, about
which Bob was completely uninformed initially, the AIG is $\mathcal{D_{\mathcal{S}}}(I_{\text{A}},I_{\text{A}},I_{\text{0}})=\mathcal{D_{\mathcal{S}}}(I_{\text{A}},I_{\text{0}})=1\,\text{bit}.$\footnote{In case Alice completely misinforms Bob about the state of a coin,
the AIG is $\mathcal{D_{\mathcal{S}}}(I_{\text{A}},I_{\text{B}},I_{\text{0}})=-\infty$.
This reflects the fact that an infinite amount of information on the
coin would be needed to correct Bob' s inappropriate world view. He
has ended up in a belief state, in which the correct situation of
head seems to him to be completely impossible. No Bayesian probability
update can get him from this state to the correct one. In such, prior
probabilities only get multiplied with finite numbers, and if one
of those prior probabilities is zero, as $P(\text{head}|I_{\text{B}})=0$,
this probability will stay zero, as the multiplication of a zero quantity
with a finite number cannot change it away from being zero.}

If one defines surprise as negative logarithmic probability (density),
$\mathcal{H}(\cdot|I):=-\ln\mathcal{P}(\cdot|I)$, then the AIG provides
Alice's expectation for Bob's surprise reduction due to her communication,

\begin{equation}
\mathcal{D_{\mathcal{S}}}(I_{\text{A}},I_{\text{B}},I_{\text{0}})=\left\langle \mathcal{H}(s|I_{\text{0}})-\mathcal{H}(s|I_{\text{B}})\right\rangle _{(s|I_{\text{A}})}.
\end{equation}
 Here, we defined expectation values according to some probability
$\mathcal{P}(s|I)$ as 
\begin{equation}
\left\langle f(s)\right\rangle _{(s|I)}:=\int\text{d}s\,f(s)\,\mathcal{P}(s|I).
\end{equation}

Note that also the in the construction of AIG involved relative entropies,
$\mathcal{D_{\mathcal{S}}}(I_{\text{A}},I_{\text{0}})$ and $\mathcal{D_{\mathcal{S}}}(I_{\text{A}},I_{\text{B}})$,
are Alice's expectations for the surprises of Bob's initial and final
state, respectively, minus her own expected surprise,
\begin{equation}
\mathcal{D_{\mathcal{S}}}(I_{\text{A}},I_{\text{0/B}})=\left\langle \mathcal{H}(s|I_{\text{0/B}})-\mathcal{H}(s|I_{\text{A}})\right\rangle _{(s|I_{\text{A}})}.
\end{equation}
Since her expectation of her own surprise does not change by the communication
act, it cancels out in the construction of the AIG, $\mathcal{D_{\mathcal{S}}}(I_{\text{A}},I_{\text{0}})-\mathcal{D_{\mathcal{S}}}(I_{\text{A}},I_{\text{B}})$
.

This allows to construct the AIG even for an absolute certain knowledge
state of Alice, $\mathcal{P}(s|I_{\text{A}})=\delta(s-m_{\text{A}})$,
a state in which her own update surprise would be infinite. The AIG
becomes then just
\begin{eqnarray}
\mathcal{D_{\mathcal{S}}}(I_{\text{A}},I_{\text{B}},I_{\text{0}}) & = & \int_{S}\text{d}s\,\delta(s-m_{\text{A}})\,\ln\frac{\mathcal{P}(s|I_{\text{B}})}{\mathcal{P}(s|I_{0})}\nonumber \\
 & = & \ln\frac{\mathcal{P}(m_{\text{A}}|I_{\text{B}})}{\mathcal{P}(m_{\text{A}}|I_{0})}\nonumber \\
 & = & \mathcal{H}(m_{\text{A}}|I_{\text{0}})-\mathcal{H}(m_{\text{A}}|I_{\text{B}}),\label{eq:ground-truth}
\end{eqnarray}
the surprise change for the value believed by Alice. This is a finite
quantity as long as none of the two involved probabilities were zero
at the location $m_{\text{A}}$.

\subsection{Memorizing}

Memorizing can be regarded as communication with the future self,
with the initial state $I_{0}$ being the ideal state $I_{\text{A}}$,
$I_{0}=I_{\text{A}}$, and the memorized state $I_{\text{B}}$ being
potentially degraded from this. This is reflected in the AIG being
never positive for $I_{0}=I_{\text{A}}$, 
\begin{eqnarray}
\mathcal{D}(I_{\text{A}},I_{\text{B}},I_{\text{0}}) & = & \mathcal{D_{\mathcal{S}}}(I_{\text{A}},I_{\text{B}},I_{\text{A}})\nonumber \\
 & = & \mathcal{D_{\mathcal{S}}}(I_{\text{A}},I_{\text{A}})-\mathcal{D_{\mathcal{S}}}(I_{\text{A}},I_{\text{B}})\nonumber \\
 & = & -\mathcal{D_{\mathcal{S}}}(I_{\text{A}},I_{\text{B}})\le0.
\end{eqnarray}
The amount of this negative gain is exactly the amount of information
lost in the act of memorization. Some Bayesian schemes to memorize
information in a compressed form aim to minimize exactly this loss
\parencites{2012arXiv1205.6935G}{harth2021toward}.

\subsection{Inference}

In inference, the situation is similar. An initial $I_{\text{0}}$,
an achieved $I_{\text{B}}$, and an ideally achieved belief state
$I_{\text{A }}$can be defined for any inference operation. The AIG
$\mathcal{D_{\mathcal{S}}}(I_{\text{A}},I_{\text{B}},I_{\text{0}})$
thereby characterizes the amount of information extracted by the act
of inference.

\subsection{Cognitive fidelity}

For the purpose of characterizing the AIG, we can treat communication,
memorization, and inference on the same footing and regard them all
as slightly differently configured information processing operations.
We summarize them in the term \emph{cognition}.

Furthermore, we can define a \textbf{cognitive fidelity} as the ratio
of the AIG to the ideal information gain in a cognitive update operation,
\begin{eqnarray}
\mathcal{E}_{\mathcal{S}}(I_{\text{A}},I_{\text{B}},I_{\text{0}}) & := & \frac{\mathcal{D_{\mathcal{S}}}(I_{\text{A}},I_{\text{B}},I_{\text{0}})}{\mathcal{D_{\mathcal{S}}}(I_{\text{A}},I_{\text{A}},I_{\text{0}})}\nonumber \\
 & = & \frac{\mathcal{D_{\mathcal{S}}}(I_{\text{A}},I_{\text{0}})-\mathcal{D_{\mathcal{S}}}(I_{\text{A}},I_{\text{B}})}{\mathcal{D_{\mathcal{S}}}(I_{\text{A}},I_{\text{0}})}\\
 & = & 1-\frac{\mathcal{D_{\mathcal{S}}}(I_{\text{A}},I_{\text{B}})}{\mathcal{D_{\mathcal{S}}}(I_{\text{A}},I_{\text{0}})}\in[-\infty,1].
\end{eqnarray}
We have $\mathcal{E}_{\mathcal{S}}(I_{\text{A}},I_{\text{A}},I_{\text{0}})=1$
for a perfect update, $\mathcal{E}_{\mathcal{S}}(I_{\text{A}},I_{\text{0}},I_{\text{0}})=0$
for no update, and $\mathcal{E}_{\mathcal{S}}(I_{\text{A}},I_{\text{B}},I_{\text{0}})<0$
for an update that goes mostly into a wrong direction, where \emph{wrong
direction }shall be defined as $\mathcal{D_{\mathcal{S}}}(I_{\text{A}},I_{\text{0}})<\mathcal{D_{\mathcal{S}}}(I_{\text{A}},I_{\text{B}})$.

\section{Relations to other measures\protect\label{sec:Relation-to-other}}

AIG is related to many of the existing information measures, like
Kullback-Leibler divergence, mutual information, Rényi divergence,
and other scoring rules to characterize probabilities and their differences.
These measures capture different aspects of probability distributions
and might therefore be ideal for different purposes, which can be
used to derive them \parencite{2024AnP...53600334E}. Before we derive
AIG axiomatically in Sec.\ \ref{sec:Axiomatic-derivation}, we discuss
its relation to a number of other information measures.

\subsection{Kullback-Leibler divergence}

The most well known information measure is the Kullback-Leibler (KL)
divergence \cite{10.1214/aoms/1177729694} aka relative entropy 
\begin{equation}
\mathcal{D}_{\text{KL}}(\mathcal{P}(s|I_{\text{A}})||\mathcal{P}(s|I_{\text{B}}))\equiv\mathcal{D_{\mathcal{S}}}(I_{\text{A}},I_{\text{B}})
\end{equation}
out of which the AIG is build as $\mathcal{D_{\mathcal{S}}}(I_{\text{A}},I_{\text{B}},I_{\text{0}})=\mathcal{D_{\mathcal{S}}}(I_{\text{A}},I_{\text{B}})-\mathcal{D_{\mathcal{S}}}(I_{\text{A}},I_{\text{0}})$.
The relative entropy is a divergence, as it is not symmetric in its
arguments, $\mathcal{D_{\mathcal{S}}}(I_{\text{A}},I_{\text{B}})\neq\mathcal{D_{\mathcal{S}}}(I_{\text{B}},I_{\text{A}})$
in general. The AIG is also not symmetric for exchanges involving
its first argument. It is, however, anti-symmetric in the other two
arguments, the ones that are involved in the information update,
\begin{eqnarray}
\mathcal{D_{\mathcal{S}}}(I_{\text{A}},I_{\text{B}},I_{\text{0}}) & \!\!\!\!=\!\!\!\! & \left\langle \ln\frac{\mathcal{P}(s|I_{\text{B}})}{\mathcal{P}(s|I_{\text{0}})}\right\rangle _{(s|I_{\text{A}})}\!\!\!\!\!\!\!\!\!\!\!=-\left\langle \ln\frac{\mathcal{P}(s|I_{\text{0}})}{\mathcal{P}(s|I_{\text{B}})}\right\rangle _{(s|I_{\text{A}})}\nonumber \\
 & \!\!\!\!=\!\!\!\! & -\mathcal{D_{\mathcal{S}}}(I_{\text{A}},I_{\text{0}},I_{\text{B}}).\label{eq:anti-symmetric}
\end{eqnarray}
This renders the AIG to be a directed distance. It measures how much
an update from $I_{\text{0}}$ to $I_{\text{B}}$ gets Bob's belief
closer to Alice's, $I_{\text{A}}.$ It is path additive in the sense,
that if Bob would continue to update to $I_{\text{C}}$, the total
AIG from $I_{\text{0}}$ to $I_{\text{C}}$ equals that from $I_{\text{0}}$
to $I_{\text{B}}$ plus that from $I_{\text{B}}$ to $I_{\text{C}}$,
\begin{eqnarray}
\mathcal{D_{\mathcal{S}}}(I_{\text{A}},I_{\text{C}},I_{\text{0}}) & \!\!\!\!=\!\!\!\! & \left\langle \ln\frac{\mathcal{P}(s|I_{\text{C}})}{\mathcal{P}(s|I_{\text{0}})}\right\rangle _{(s|I_{\text{A}})}\nonumber \\
 & \!\!\!\!=\!\!\!\! & \left\langle \ln\frac{\mathcal{P}(s|I_{\text{B}})\,\mathcal{P}(s|I_{\text{C}})}{\mathcal{P}(s|I_{\text{B}})\,\mathcal{P}(s|I_{\text{0}})}\right\rangle _{(s|I_{\text{A}})}\nonumber \\
 & \!\!\!\!=\!\!\!\! & \left\langle \ln\frac{\mathcal{P}(s|I_{\text{C}})}{\mathcal{P}(s|I_{\text{B}})}+\ln\frac{\mathcal{P}(s|I_{\text{B}})}{\mathcal{P}(s|I_{\text{0}})}\right\rangle _{(s|I_{\text{A}})}\nonumber \\
 & \!\!\!\!=\!\!\!\! & \mathcal{D_{\mathcal{S}}}(I_{\text{A}},I_{\text{C}},I_{\text{B}})+\mathcal{D_{\mathcal{S}}}(I_{\text{A}},I_{\text{B}},I_{\text{0}}).
\end{eqnarray}
The KL is not path additive in general, as 
\begin{eqnarray}
\mathcal{D_{\mathcal{S}}}(I_{\text{A}},I_{\text{C}}) & \!\!\!\!=\!\!\!\! & \left\langle \ln\left[\frac{\mathcal{P}(s|I_{\text{A}})}{\mathcal{P}(s|I_{\text{B}})}\frac{\mathcal{P}(s|I_{\text{B}})}{\mathcal{P}(s|I_{\text{C}})}\right]\right\rangle _{(s|I_{\text{A}})}\nonumber \\
 & {\color{red}\!\!\!\!\neq\!\!\!\!} & \left\langle \ln\left[\frac{\mathcal{P}(s|I_{\text{A}})}{\mathcal{P}(s|I_{\text{B}})}\left(\frac{\mathcal{P}(s|I_{\text{B}})}{\mathcal{P}(s|I_{\text{C}})}\right)^{\frac{\mathcal{P}(s|I_{\text{B}})}{\mathcal{P}(s|I_{\text{A}})}}\right]\right\rangle _{(s|I_{\text{A}})}\nonumber \\
 & \!\!\!\!=\!\!\!\! & \left\langle \ln\frac{\mathcal{P}(s|I_{\text{A}})}{\mathcal{P}(s|I_{\text{B}})}+\frac{\mathcal{P}(s|I_{\text{B}})}{\mathcal{P}(s|I_{\text{A}})}\ln\frac{\mathcal{P}(s|I_{\text{B}})}{\mathcal{P}(s|I_{\text{C}})}\right\rangle _{(s|I_{\text{A}})}\nonumber \\
 & \!\!\!\!=\!\!\!\! & \left\langle \ln\frac{\mathcal{P}(s|I_{\text{A}})}{\mathcal{P}(s|I_{\text{B}})}\right\rangle _{(s|I_{\text{A}})}+\left\langle \ln\frac{\mathcal{P}(s|I_{\text{B}})}{\mathcal{P}(s|I_{\text{C}})}\right\rangle _{(s|I_{\text{B}})}\nonumber \\
 & \!\!\!\!=\!\!\!\! & \mathcal{D_{\mathcal{S}}}(I_{\text{A}},I_{\text{B}})+\mathcal{D_{\mathcal{S}}}(I_{\text{B}},I_{\text{C}})\label{eq:path-additive}
\end{eqnarray}
for $\mathcal{P}(s|I_{\text{B}})\neq\mathcal{P}(s|I_{\text{A}})$.

Finally, AIG reduces to the KL divergence in case $I_{\text{A}}=I_{\text{B}},$
\begin{equation}
\mathcal{D_{\mathcal{S}}}(I_{\text{A}},I_{\text{A}},I_{\text{0}})=\left\langle \ln\frac{\mathcal{P}(s|I_{\text{A}})}{\mathcal{P}(s|I_{\text{0}})}\right\rangle _{(s|I_{\text{A}})}=\mathcal{D_{\mathcal{S}}}(I_{\text{A}},I_{\text{0}})\label{eq:AIG_reduces_to_KL}
\end{equation}

\subsection{Mutual information}

Mutual information (MI) \parencites{gel1957computation}{duncan1970calculation}
measures how much the knowledge $I_{\text{B}}$ on a situation $s=(x,y)$
that consists out of two unknowns $x\in\mathcal{X}$ and $y\in\mathcal{Y}$
with $\mathcal{S}=\mathcal{X}\times\mathcal{Y}$ has more information
than knowing only their marginals,
\begin{eqnarray}
\text{MI}_{\mathcal{X},\mathcal{Y}}(I_{\text{B}}) & := & \left\langle \ln\frac{\mathcal{P}(x,y|I_{\text{B}})}{\mathcal{P}(x|I_{\text{B}})\mathcal{\,P}(y|I_{\text{B}})}\right\rangle _{(x,y|I_{\text{B}})}.\nonumber \\
 & = & \mathcal{D_{\mathcal{S}}}(I_{\text{B}},I_{\text{0}}),\text{ with}\\
\mathcal{P}(x,y|I_{\text{0}}) & := & \mathcal{P}(x|I_{\text{B}})\mathcal{\,P}(y|I_{\text{B}})
\end{eqnarray}
the knowledge state ignoring correlations between the variables.

It is as well possible to introduce an achieved mutual information
(AMI) from the perspective of Alice by defining

\begin{eqnarray}
\text{AMI}_{\mathcal{X},\mathcal{Y}}(I_{\text{A}},I_{\text{B}}) & := & \left\langle \ln\frac{\mathcal{P}(x,y|I_{\text{B}})}{\mathcal{P}(x|I_{\text{B}})\mathcal{\,P}(y|I_{\text{B}})}\right\rangle _{(x,y|I_{\text{A}})}\nonumber \\
 & = & \mathcal{D_{\mathcal{S}}}(I_{\text{A}},I_{\text{B}},I_{\text{0}})\text{, again with}\\
\mathcal{P}(x,y|I_{\text{0}}) & := & \mathcal{P}(x|I_{\text{B}})\mathcal{\,P}(y|I_{\text{B}}).\nonumber 
\end{eqnarray}
Different to MI, which is always positive or zero, AMI can become
negative. For example, in case $I_{\text{A}}=I_{\text{0}}$, meaning
that Alice is sure that the correlations Bob believes in are spurious,
but thinks that his marginal distribution are correct, then $\text{AMI}_{\mathcal{X},\mathcal{Y}}(I_{\text{A}},I_{\text{B}})=\mathcal{D_{\mathcal{S}}}(I_{\text{0}},I_{\text{B}},I_{\text{0}})=-\mathcal{D_{\mathcal{S}}}(I_{\text{0}},I_{\text{0}},I_{\text{B}})=-\mathcal{D_{\mathcal{S}}}(I_{\text{0}},I_{\text{B}})\le0$.
This holds as the AIG is anti-symmetric, Eq.\ \ref{eq:anti-symmetric},
reduces to the KL if the first arguments coincide, Eq.\ \ref{eq:AIG_reduces_to_KL},
and the KL divergence is never negative, Eq.\ \ref{eq:KL_is_positive}.

\subsection{Rényi divergence}

The Rényi or alpha-divergence \parencite{renyi1961measures} can be
defined as
\begin{eqnarray}
\mathcal{D_{\mathcal{S}}^{\alpha}}(I_{\text{A}},I_{\text{0}}) & \!\!\!\!=\!\!\!\! & \ln\left\langle \left(\frac{\mathcal{P}(s|I_{\text{A}})}{\mathcal{P}(s|I_{\text{0}})}\right)^{\alpha-1}\right\rangle _{(s|I_{\text{A}})}^{\frac{1}{\alpha-1}},
\end{eqnarray}
which has the natural generalization to become an achieved alpha-information
gain ($\alpha$-AIG) via 
\begin{eqnarray}
\mathcal{D_{\mathcal{S}}^{\alpha}}(I_{\text{A}},I_{\text{B}},I_{\text{0}}) & \!\!\!\!=\!\!\!\! & \ln\left\langle \left(\frac{\mathcal{P}(s|I_{\text{B}})}{\mathcal{P}(s|I_{\text{0}})}\right)^{\alpha-1}\right\rangle _{(s|I_{\text{A}})}^{\frac{1}{\alpha-1}}.
\end{eqnarray}
$\alpha$-AIG reduces to AIG in the limit of $\alpha\rightarrow1$.
Different to AIG, $\alpha$-AIG is in general not anti-symmetric in
its last two arguments (Eq.\ \ref{eq:anti-symmetric}), nor path
additive (Eq.\ \ref{eq:path-additive}).

\subsection{Scoring rules}

A scoring rule $\mathbb{S}(\mathcal{P},s)$ quantifies the performance
of a probability, for example $\mathcal{P}(s)=\mathcal{P}(s|I_{\text{B}})$,
in predicting $s\in\mathcal{S}$ by a real number \parencite{selten1998axiomatic,gneiting2007strictly,landes2015probabilism}.
Its score under knowledge $I_{\text{A}}$ is the expectation
\begin{equation}
\mathcal{D}_{\mathcal{S}}^{\mathbb{S}}(I_{\text{A}},I_{\text{B}}):=\left\langle \mathbb{S}(\mathcal{P}(\cdot|I_{\text{B}}),s)\right\rangle _{(s|I_{\text{A}})}.
\end{equation}
Examples of scoring rules include $\mathbb{S_{\text{CE}}}(\mathcal{P},s)=\ln\mathcal{P}(s)$,
which leads to cross-entropy
\begin{equation}
\mathcal{D}_{\mathcal{S}}^{\mathbb{S}_{\text{CE}}}(I_{\text{A}},I_{\text{B}})=\left\langle \ln(\mathcal{P}(s|I_{\text{B}}))\right\rangle _{(s|I_{\text{A}})},
\end{equation}
$\mathbb{S_{\text{RE}}}(\mathcal{P},s)=-\ln\left[\mathcal{P}(s)/\mathcal{P}(s|I_{\text{A}})\right]$,
that reproduces the relative entropy

\begin{equation}
\mathcal{D}_{\mathcal{S}}^{\mathbb{S_{\text{RE}}}}(I_{\text{A}},I_{\text{B}})=\left\langle \ln\frac{\mathcal{P}(s|I_{\text{A}})}{\mathcal{P}(s|I_{\text{B}})}\right\rangle _{(s|I_{\text{A}})}=\mathcal{D_{\mathcal{S}}}(I_{\text{A}},I_{\text{B}}),
\end{equation}
$\mathbb{S_{\text{AIG}}}(\mathcal{P},s)=\ln\left[\mathcal{P}(s)/\mathcal{P}(s|I_{\text{0}})\right]$,
which generates AIG

\begin{equation}
\mathcal{D}_{\mathcal{S}}^{\mathbb{S_{\text{AIG}}}}(I_{\text{A}},I_{\text{B}})=\left\langle \ln\frac{\mathcal{P}(s|I_{\text{B}})}{\mathcal{P}(s|I_{\text{0}})}\right\rangle _{(s|I_{\text{A}})}=\mathcal{D_{\mathcal{S}}}(I_{\text{A}},I_{\text{B}},I_{\text{0}}),
\end{equation}
and $\mathbb{S_{\alpha\text{-AIG}}}(\mathcal{P},s)=\left[\mathcal{P}(s)/\mathcal{P}(s|I_{\text{0}})\right]^{\alpha-1},$which
generates the central element of $\alpha$-AIG
\begin{eqnarray}
\mathcal{D}_{\mathcal{S}}^{\mathbb{S_{\alpha\text{-AIG}}}}(I_{\text{A}},I_{\text{B}}) & = & \left\langle \left(\frac{\mathcal{P}(s|I_{\text{B}})}{\mathcal{P}(s|I_{\text{0}})}\right)^{\alpha-1}\right\rangle _{(s|I_{\text{A}})}\nonumber \\
 & = & \exp\left[(\alpha-1)\,\mathcal{D_{\mathcal{S}}^{\alpha}}(I_{\text{A}},I_{\text{B}},I_{\text{0}})\right].
\end{eqnarray}
All these scoring rules are local, meaning $\mathbb{S}(\mathcal{P},s)=f(\mathcal{P}(s),s)$
with $f:\mathbb{R}\times\mathcal{S}\rightarrow\mathbb{R}$ being only
sensitive to the value of $\mathcal{P}$ at location $s$, and not
to those at other locations. These rules are also all proper, in the
sense that they are at a minimum or maximum for $\mathcal{P}(s)\equiv\mathcal{P}(s|I_{\text{A}})$
and only for this case. Thus, extremizing them for $I_{\text{B}}$
yields $I_{\text{B}}=I_{\text{A}}$.

\subsection{Information geometry}

The differential analysis of relative entropy has led to the rich
field of information geometry \parencite{amari1997information,ay2017information}.
The key observation is that a Taylor expansion in $\Delta:=I_{\text{B}}-I_{\text{A}}$
up to second order
\begin{eqnarray}
\mathcal{D}_{\mathcal{S}}(I_{\text{A}},I_{\text{B}}) & \!\!\!\!=\!\!\!\! & \overbrace{\mathcal{D}_{\mathcal{S}}(I_{\text{A}},I_{\text{A}})}^{=0}+\overbrace{\left(\frac{\partial\mathcal{D}_{\mathcal{S}}(I_{\text{A}},I_{\text{B}})}{\partial I_{\text{B}}}\right)_{I_{\text{B}}=I_{\text{A}}}^{\text{t}}}^{=0}\!\!\!\!\!\!\!\!\!\!\!\!\Delta\nonumber \\
 &  & +\frac{1}{2}\Delta^{t}\left.\frac{\partial^{2}\mathcal{D}_{\mathcal{S}}(I_{\text{A}},I_{\text{B}})}{\partial I_{\text{B}}\partial I_{\text{B}}^{\text{t}}}\right|_{I_{\text{B}}=I_{\text{A}}}\!\!\!\!\!\!\!\!\!\!\!\!\Delta+\mathcal{O}(\Delta^{3})\nonumber \\
 & \!\!\!\!=\!\!\!\! & \frac{1}{2}\Delta^{t}g_{I_{\text{A}}}\Delta+\mathcal{O}(\Delta^{3})
\end{eqnarray}
leads to an expression that seems to contain a metric as can be found
in differential geometry, 
\begin{eqnarray}
g_{I_{\text{A}}} & := & \left.\frac{\partial^{2}\mathcal{D}_{\mathcal{S}}(I_{\text{A}},I_{\text{B}})}{\partial I_{\text{B}}\partial I_{\text{B}}^{\text{t}}}\right|_{I_{\text{B}}=I_{\text{A}}}\nonumber \\
 & = & \left.\left\langle \frac{\partial^{2}\mathcal{H}(s|I_{\text{B}})}{\partial I_{\text{B}}\partial I_{\text{B}}^{\text{t}}}\right\rangle _{(s|I_{\text{A}})}\right|_{I_{\text{B}}=I_{\text{A}}}\nonumber \\
 & = & \left\langle \frac{\partial^{2}\mathcal{H}(s|I_{\text{A}})}{\partial I_{\text{A}}\partial I_{\text{A}}^{\text{t}}}\right\rangle _{(s|I_{\text{A}})}\nonumber \\
 & = & \left\langle \frac{\partial}{\partial I_{\text{A}}}\frac{-1}{\mathcal{P}(s|I_{\text{A}})}\frac{\partial\mathcal{P}(s|I_{\text{A}})}{\partial I_{\text{A}}^{\text{t}}}\right\rangle _{(s|I_{\text{A}})}\nonumber \\
 & = & \left\langle \frac{1}{\mathcal{P}^{2}(s|I_{\text{A}})}\frac{\partial\mathcal{P}(s|I_{\text{A}})}{\partial I_{\text{A}}}\frac{\partial\mathcal{P}(s|I_{\text{A}})^{\text{t}}}{\partial I_{\text{A}}}\right\rangle _{(s|I_{\text{A}})}\nonumber \\
 &  & -\underbrace{\left\langle \frac{1}{\mathcal{P}(s|I_{\text{A}})}\frac{\partial^{2}\mathcal{P}(s|I_{\text{A}})}{\partial I_{\text{A}}\partial I_{\text{A}}^{\text{t}}}\right\rangle _{(s|I_{\text{A}})}}_{=\partial_{I_{\text{A}}}^{2}\int_{\mathcal{S}}ds\,\mathcal{P}(s|I_{\text{A}})=\partial_{I_{\text{A}}}^{2}1=0}\nonumber \\
 & = & \left\langle \frac{\partial\mathcal{H}(s|I_{\text{A}})}{\partial I_{\text{A}}}\frac{\partial\mathcal{H}(s|I_{\text{A}})}{\partial I_{\text{A}}}^{\text{t}}\right\rangle _{(s|I_{\text{A}})}.
\end{eqnarray}
In particular the last expression makes clear that the ``information
metric'' is positive definite, $g_{I_{\text{A}}}\ge0$. As $g_{I_{\text{A}}}$
depends on the location $I_{\text{A}}$ in the knowledge space, it
seems to specify a Riemann manifold.

$\mathcal{D}_{\mathcal{S}}(I_{\text{A}},I_{\text{B}},I_{\text{0}})=\mathcal{D}_{\mathcal{S}}(I_{\text{A}},I_{\text{0}})-\mathcal{D}_{\mathcal{S}}(I_{\text{A}},I_{\text{B}})$
can be expanded in a similar way around the ideal knowledge state
$I_{\text{B}}=I_{\text{A}}+\Delta$, yielding 
\begin{equation}
\mathcal{D}_{\mathcal{S}}(I_{\text{A}},I_{\text{B}},I_{\text{0}})=\mathcal{D}_{\mathcal{S}}(I_{\text{A}},I_{\text{0}})-\frac{1}{2}\Delta^{t}g_{I_{\text{A}}}\Delta+\mathcal{O}(\Delta^{3})
\end{equation}
with the same metric $g_{I_{\text{A}}}$.

Expanding the AIG around the initial knowledge state $I_{\text{B}}=I_{\text{0}}+\Delta$,
however, gives a non zero linear term, 
\begin{eqnarray}
\mathcal{D}_{\mathcal{S}}(I_{\text{A}},I_{\text{B}},I_{\text{0}}) & \!\!\!\!=\!\!\!\! & \mathcal{D}_{\mathcal{S}}(I_{\text{A}},I_{\text{0}},I_{\text{0}})+\left.\frac{\partial\mathcal{D}_{\mathcal{S}}(I_{\text{A}},I_{\text{B}},I_{\text{0}})}{\partial I_{\text{B}}}\right|_{I_{\text{B}}=I_{\text{0}}}^{\text{t}}\!\!\!\!\!\!\!\!\!\!\!\!\Delta\nonumber \\
 &  & +\frac{1}{2}\Delta^{t}\left.\frac{\partial^{2}\mathcal{D}_{\mathcal{S}}(I_{\text{A}},I_{\text{B}},I_{\text{0}})}{\partial I_{\text{B}}\partial I_{\text{B}}^{\text{t}}}\right|_{I_{\text{B}}=I_{\text{0}}}\!\!\!\!\!\!\!\!\!\!\!\!\Delta+\mathcal{O}(\Delta^{3})\nonumber \\
 & \!\!\!\!=\!\!\!\! & \mathcal{D}_{\mathcal{S}}(I_{\text{A}},I_{\text{0}})-j_{I_{\text{0}}}^{\text{t}}\Delta-\frac{1}{2}\Delta^{t}f_{I_{\text{0}}}\Delta+\mathcal{O}(\Delta^{3})\text{,}\nonumber \\
\\\text{ with }j_{I_{\text{0}}} & \!\!\!\!=\!\!\!\! & \left\langle \frac{\partial\mathcal{H}(s|I_{\text{0}})}{\partial I_{\text{0}}}\right\rangle _{(s|I_{\text{A}})}\text{ and}\\
f_{I_{\text{0}}} & \!\!\!\!=\!\!\!\! & g_{I_{\text{0}}}-\int_{\mathcal{S}}ds\frac{\mathcal{P}(s|I_{\text{A}})}{\mathcal{P}(s|I_{\text{0}})}\frac{\partial^{2}\mathcal{P}(s|I_{\text{0}})}{\partial I_{\text{0}}\partial I_{\text{0}}^{\text{t}}}.
\end{eqnarray}
Although $g_{I_{\text{0}}}$ is still a positive definite matrix,
the Hessian $f_{I_{\text{0}}}$ is not necessarily positive definite
any more, as the term with the second order derivative does not need
to vanish. It is therefore less natural to interpret it as a metric
of a Riemann manifold.

Nevertheless, the positive definite part of the Hessian, $g_{I_{\text{0}}}$,
allows to construct AIG optimizing Newton schemes that update $I_{0}$
to states with a larger AIG by going in the direction given by $-g_{I_{\text{0}}}^{-1}j_{I_{0}}$.
Such updates are used to solve ultra high dimensional inference problems
via Metric Gaussian Variational Inference \parencite{knollmuller2019metric}
and geometric Variational Inference \parencite{2021Entrp..23..853F}.

\section{Axiomatic derivation\protect\label{sec:Axiomatic-derivation}}

\subsection{Axioms}

Our axiomatic derivation of AIG follows closely that of relative entropy
as given in \parencite{2024AnP...53600334E}, which itself followed
and extended previous works \cite{10.1214/aoms/1177729694,2017Entrp..19..402L}.
As before, we assume that Alice, who has has superior knowledge $I_{\text{A}}$
about an unknown quantity $s\in\mathcal{S}$ compared to Bob, sends
him a message that lets him change his initial knowledge $I_{\text{0}}$
to an updated one $I_{\text{B}}$. We ask for a measure of success
that allows Alice to decide which message she should send to Bob.
This measure should follow from principles or axioms.

Alice believes that her knowledge $I_{\text{A}}$ on $s$ is better
than that of Bob, which she knows to be initially $I_{\text{0}}$
and which she can predict to be $I_{\text{B}}$ after his update induced
by her message. Thus, basically she can chose $I_{\text{B}}$ out
of a set of knowledge states that she can address with her communication.
She wants to have a measure that leads this decision and this measure
should obey a number of desirable axioms. Although this measure will
turn out to be the AIG up to a multiplicative factor, for the moment,
we call it just the gain $G_{\mathcal{S}}(I_{\text{A}},I_{\text{B}},I_{\text{0}})$.

Our axiomatic requirements are that the gain should be \emph{additive}
w.r.t.\ to Alice's belief on the possible values of $s$, that it
is \emph{analytical} with w.r.t.\ to the relevant argument, Bob's
final knowledge $\mathcal{P}(s|I_{\text{B}})$, \emph{proper} w.r.t.\ to
her knowledge, meaning that it is maximal for the proper update $I_{\text{B}}=I_{\text{A}}$,
and \emph{calibrated} such that it vanishes for no update by Bob,
$I_{\text{B}}=I_{\text{0}}$. In more detail, these requirements are:
\begin{description}
\item [{\textcolor{black}{Additive:}}] \textcolor{black}{In case Alice
knowledge $I_{\text{A}}$ is hierarchical, with $\mathcal{P}(s|I_{\text{A}})=\sum_{i=1}^{n}p_{\text{A,}i}\,\mathcal{P}(s|I_{\text{A},i})$
and $I_{\text{A}}=(I_{\text{A},i},p_{i})_{i=1}^{n}$ being possible
knowledge sub-states $I_{\text{A},i}$ for Alice and the by her associated
probabilities $p_{i}$, the gain calculated for the perspective of
$I_{\text{A}}$ should be the the sum of the gains $G_{\mathcal{S}}(I_{\text{A,}i},I_{\text{B}},I_{\text{0}})$
of the sub-states $I_{\text{A,}i},$ weighted by their corresponding
probabilities $p_{\text{A,}i}$: $G_{\mathcal{S}}(I_{\text{A}},I_{\text{B}},I_{\text{0}})=\sum_{i=1}^{n}p_{\text{A,}i}\,G_{\mathcal{S}}(I_{\text{A},i},I_{\text{B}},I_{\text{0}})$.
This also holds for a continuous set of possible knowledge states
$I_{\text{A}}(x)$ with $p_{\text{A}}(x)$ and $x\in\mathcal{X}$,
such that $G_{\mathcal{S}}(I_{\text{A}},I_{\text{B}},I_{\text{0}})=\int_{\mathcal{X}}\text{d}x\,p_{\text{A}}(x)\,G_{\mathcal{S}}(I_{\text{A}}(x),I_{\text{B}},I_{\text{0}})$.}
\item [{\textcolor{black}{Locality:}}] \textcolor{black}{In case Alice
knows which situation $s'$ will happen, $I_{\text{A}}=I_{s=s'}$,
with $I_{s=s'}$ expressing certainty that $s=s'$, $\mathcal{P}(s|I_{s=s'}):=\delta(s-s')$,}\footnote{This is a slight abuse of notation, as $I_{s=s'}$ is not necessarily
in $\mathcal{I}$. For some distributions, $I_{s=s'}$ is given by
a limit within $\mathcal{I}$, e.g.\ in case of a multivariate Gaussian,
by the limit of vanishing covariance.}\textcolor{black}{{} the gain should only depend on the probability
Bob finally assigns to this case, $\mathcal{P}(s'|I_{\text{B}})$,
but not on any probability he assigns to other cases, $\mathcal{P}(s''|I_{\text{B}})$
for $s''\neq s'$.}
\item [{\textcolor{black}{Analytical:}}] \textcolor{black}{Since Alice
needs to maximize the gain w.r.t. to $I_{\text{B }}$, the gain should
be infinitely differentiable w.r.t.\ $\mathcal{P}(s|I_{\text{B}})$
at all viable values of $I_{\text{B}}$. This means it must be analytical.}
\item [{Proper:}] In case Alice is able to send a message to Bob that leads
to $I_{\text{B}}=I_{\text{A}}$, the gain should tell her to do so,
implying $G_{\mathcal{S}}(I_{\text{A}},I_{\text{B}},I_{\text{0}})$
is maximal for $I_{\text{B}}=I_{\text{A}}$.
\item [{Calibration:}] The gain should vanish in case Bob does not update,
implying $G_{\mathcal{S}}(I_{\text{A}},I_{\text{0}},I_{\text{0}})=0$.
\end{description}
In comparison to \parencite{2024AnP...53600334E}, the \emph{additive}
axiom has been introduced here. Furthermore, the chosen calibration
differs. The latter was modified here since we are interested in a
gain in moving Bob's knowledge away from $I_{\text{0}}$, and in \parencite{2024AnP...53600334E}
the perspective was on a loss w.r.t.\ the optimal update $I_{\text{B}}=I_{\text{A}}$.
The derivation of AIG, however, continues very analogously.

We decompose $I_{\text{A}}$ into atomic believes over the set $\mathcal{S}$
by identifying $x=s$ and $\mathcal{X}=\mathcal{S}$, such that $p_{\text{A}}(x):=\mathcal{P}(s=x|I_{\text{A}})$
and $I_{\text{A}}(x):=I_{s=x}$, and find -- since the gain is \emph{additive}
-- that
\begin{eqnarray}
G_{\mathcal{S}}(I_{\text{A}},I_{\text{B}},I_{\text{0}}) & = & \int_{\mathcal{X}}\text{d}x\,p_{\text{A}}(x)\,G_{\mathcal{S}}(I_{\text{A}}(x),I_{\text{B}},I_{\text{0}})\nonumber \\
 & = & \int_{\mathcal{S}}\text{d}s'\,\mathcal{P}(s'|I_{\text{A}})\,G_{\mathcal{S}}(I_{s=s'},I_{\text{B}},I_{\text{0}})\nonumber \\
 & = & \langle G_{\mathcal{S}}(I_{s=s'},I_{\text{B}},I_{\text{0}})\rangle_{(s'|I_{\text{A}})}.
\end{eqnarray}

Locality implies then that 
\begin{eqnarray}
G_{\mathcal{S}}(I_{s=s'},I_{\text{B}},I_{\text{0}}) & = & g(s',\mathcal{P}(s'|I_{\text{B}}),I_{\text{0}})\nonumber \\
 &  & +\lambda\,\left(1-\int_{\mathcal{S}}\text{d}s''\,\mathcal{P}(s''|I_{\text{B}})\right),
\end{eqnarray}
with $g(s,q(s),I_{\text{0}})$ a function to be determined, and with
$\lambda$ an arbitrary Lagrange multiplier ensuring that Bob's final
knowledge states is properly normalized. This allows us to perform
the maximization of $G_{\mathcal{S}}(I_{\text{A}},I_{\text{B}},I_{\text{0}})$
by taking functional derivatives w.r.t.\ to the values of $q(s)=\mathcal{P}(s|I_{\text{B}})$
and $\lambda.$ Properness requires that the gain should be maximal
for $q(s)=\mathcal{P}(s|I_{\text{A}})$ and thus

\begin{eqnarray}
\!\!\!\!\!\!\! & \!\!\!\!\!\!\! & 0=\left.\frac{\delta G_{\mathcal{S}}(I_{\text{A}},I_{\text{B}},I_{\text{0}})}{\delta\mathcal{P}(s'|I_{\text{B}})}\right|_{I_{\text{B}}=I_{\text{A}}}\!\!\!\!\!=\nonumber \\
\!\!\!\!\!\!\! & \!\!\!\!\!\!\! & \left.\frac{\delta}{\delta q(s')}\left\langle g(s,q(s),I_{\text{0}})+\!\left(\!1\!-\!\!\int_{\mathcal{S}}\!\!\!\text{d}s''\,q(s'')\right)\lambda\right\rangle _{\!(s|I_{\text{A}})}\right|_{\!q=\mathcal{P}(\cdot|I_{\text{A}})}\!\!\!\!\!\!\!=\nonumber \\
\!\!\!\!\!\!\! & \!\!\!\!\!\!\! & \left.\frac{\delta}{\delta q(s')}\left[\left\langle g(s,q(s),I_{\text{0}})\right\rangle _{\!(s|I_{\text{A}})}\!\!+\!\left(\!1\!-\!\!\int_{\mathcal{S}}\!\!\!\text{d}s''\,q(s'')\right)\lambda\right]\right|_{\!q=\mathcal{P}(\cdot|I_{\text{A}})}\!\!\!\!\!\!\!=\nonumber \\
\!\!\!\!\!\!\! & \!\!\!\!\!\!\! & \left.\frac{\partial g(s',q(s'),I_{\text{0}})}{\partial q(s')}\,\mathcal{P}(s'|I_{\text{A}})-\lambda\right|_{q=\mathcal{P}(\cdot|I_{\text{A}})}\!\!\!\!\!\!\!=\nonumber \\
\!\!\!\!\!\!\! & \!\!\!\!\!\!\! & \left.\frac{\partial g(s',\tilde{q},I_{\text{0}})}{\partial\tilde{q}}\,\tilde{q}-\lambda\right|_{\tilde{q}=\mathcal{P}(s'|I_{\text{A}})}.
\end{eqnarray}
From this ordinary differential equation it follows that
\begin{equation}
g(s,\tilde{q},I_{\text{0}})=\lambda\,\ln\tilde{q}+c(s,I_{\text{0}}),
\end{equation}
where $c(s,I_{\text{0}})$ is to be determined via the calibration.
This is the condition
\begin{eqnarray}
0 & = & G_{\mathcal{S}}(I_{\text{A}},I_{\text{0}},I_{\text{0}})\nonumber \\
 & = & \left\langle g(s,\mathcal{P}(s|I_{\text{0}}),I_{\text{0}})\right\rangle _{(s|I_{\text{A}})}\nonumber \\
 & = & \left\langle \lambda\,\ln\mathcal{P}(s|I_{\text{0}})+c(s,I_{\text{0}})\right\rangle _{(s|I_{\text{A}})}.
\end{eqnarray}
Here we used the normalization $\int_{\mathcal{S}}\text{d}s\,\mathcal{P}(s|I_{\text{0}})=1$,
which follows from extremizing w.r.t.\ $\lambda$. This implies
\begin{equation}
\left\langle c(s,I_{\text{0}})\right\rangle _{(s|I_{\text{A}})}=-\lambda\,\left\langle \ln\mathcal{P}(s|I_{\text{0}})\right\rangle _{(s|I_{\text{A}})},
\end{equation}
which is specific enough to determine the gain 
\begin{eqnarray}
G_{\mathcal{S}}(I_{\text{A}},I_{\text{B}},I_{\text{0}}) & = & \left\langle g(s,\mathcal{P}(s|I_{\text{B}}),I_{\text{0}})\right\rangle _{(s|I_{\text{A}})}\nonumber \\
 & = & \left\langle \lambda\,\ln\mathcal{P}(s|I_{\text{B}})+c(s,I_{\text{0}})\right\rangle _{(s|I_{\text{A}})}\nonumber \\
 & = & \left\langle \lambda\,\ln\mathcal{P}(s|I_{\text{B}})-\lambda\,\ln\mathcal{P}(s|I_{\text{0}})\right\rangle _{(s|I_{\text{A}})}\nonumber \\
 & = & \lambda\,\mathcal{D_{\mathcal{S}}}(I_{\text{A}},I_{\text{B}},I_{\text{0}}).
\end{eqnarray}
Thus, the gain is the AIG times some arbitrary factor $\lambda>0$
that determines the units. This relation would only hold in a infinitesimal
environment of $\mathcal{P}(s|I_{\text{B}})\equiv\mathcal{P}(s|I_{\text{A}})$,
for which our calculation holds, but the requirement of the gain being
analytical extends it to the full domain of probability densities.
This completes the axiomatic derivation.

\subsection{Separability}

A nice property of this derivation is that the property of AIG to
be additive in case of \textbf{separability} of the involved probabilities,
which is often requested in derivations of entropy, emerges here as
a consequence. To see this, we assume that $s=(s_{1}^{\text{t}},s_{2}^{\text{t}})^{\mathrm{t}}$
can be split into two parts, with $s_{1}\in\mathcal{S}_{1}$, $s_{2}\in\mathcal{S}_{2}$,
and $\mathcal{S}=\mathcal{S}_{1}\otimes\mathcal{S}_{2}$ and that
Bob's initial and final states are separable, $\mathcal{P}(s|I_{\text{0}})=\mathcal{P}(s_{1}|I_{\text{0}})\,\mathcal{P}(s_{2}|I_{\text{0}})$,
$\mathcal{P}(s|I_{\text{B}})=\mathcal{P}(s_{1}|I_{\text{B}})\,\mathcal{P}(s_{2}|I_{\text{B}}),$
and $s^{\text{t}}$ denoting transposition of a vector or matrix.
Then we find
\begin{eqnarray}
\mathcal{D}_{\mathcal{S}}(I_{\text{A}},I_{\text{B}},I_{\text{0}}) & \!\!\!\!\!=\!\!\!\!\! & \left\langle \ln\frac{\mathcal{P}(s_{1}|I_{\text{B}})\,\mathcal{P}(s_{2}|I_{\text{B}})}{\mathcal{P}(s_{1}|I_{\text{0}})\,\mathcal{P}(s_{2}|I_{\text{0}})}\right\rangle _{(s|I_{\text{A}})}\nonumber \\
 & \!\!\!\!\!=\!\!\!\!\! & \left\langle \ln\frac{\mathcal{P}(s_{1}|I_{\text{B}})}{\mathcal{P}(s_{1}|I_{\text{0}})}\right\rangle _{(s|I_{\text{A}})}\!\!\!\!\!\!\!+\left\langle \ln\frac{\mathcal{P}(s_{2}|I_{\text{B}})}{\mathcal{P}(s_{2}|I_{\text{0}})}\right\rangle _{(s|I_{\text{A}})}\nonumber \\
 & \!\!\!\!\!=\!\!\!\!\! & \left\langle \ln\frac{\mathcal{P}(s_{1}|I_{\text{B}})}{\mathcal{P}(s_{1}|I_{\text{0}})}\right\rangle _{(s_{1}|I_{\text{A}})}\!\!\!\!\!\!\!+\left\langle \ln\frac{\mathcal{P}(s_{2}|I_{\text{B}})}{\mathcal{P}(s_{2}|I_{\text{0}})}\right\rangle _{(s_{2}|I_{\text{A}})}\nonumber \\
 & \!\!\!\!\!=\!\!\!\!\! & \mathcal{D}_{\mathcal{S}_{1}}(I_{\text{A}},I_{\text{B}},I_{\text{0}})+\mathcal{D}_{\mathcal{S}_{2}}(I_{\text{A}},I_{\text{B}},I_{\text{0}}).
\end{eqnarray}
It is an interesting observation that this holds even in case Alice's
knowledge state is not separable (see e.g. Sect.\ \ref{subsec:Mean-Field-Approximation}),
in which Alice knows about correlations between $s_{1}$ and $s_{\text{2}},$
$\mathcal{P}(s|I_{\text{A}})\neq\mathcal{P}(s_{1}|I_{\text{A}})\,\mathcal{P}(s_{2}|I_{\text{A}})$.
In this situation, the relative entropies $\mathcal{D}_{\mathcal{S}}(I_{\text{A}},I_{\text{0}})$
and $\mathcal{D}_{\mathcal{S}}(I_{\text{A}},I_{\text{B}})$, out of
which $\mathcal{D}_{\mathcal{S}}(I_{\text{A}},I_{\text{B}},I_{\text{0}})$
is composed, are both not separable themselves. However, as Bob has
no notion of the correlation, neither before nor after the update,
his AIG is fully separable for the two variables.

\section{Instructive examples\protect\label{sec:Instructive-examples}}

\begin{figure*}
\includegraphics[viewport=45bp 20bp 805bp 510bp,clip,width=0.49\textwidth]{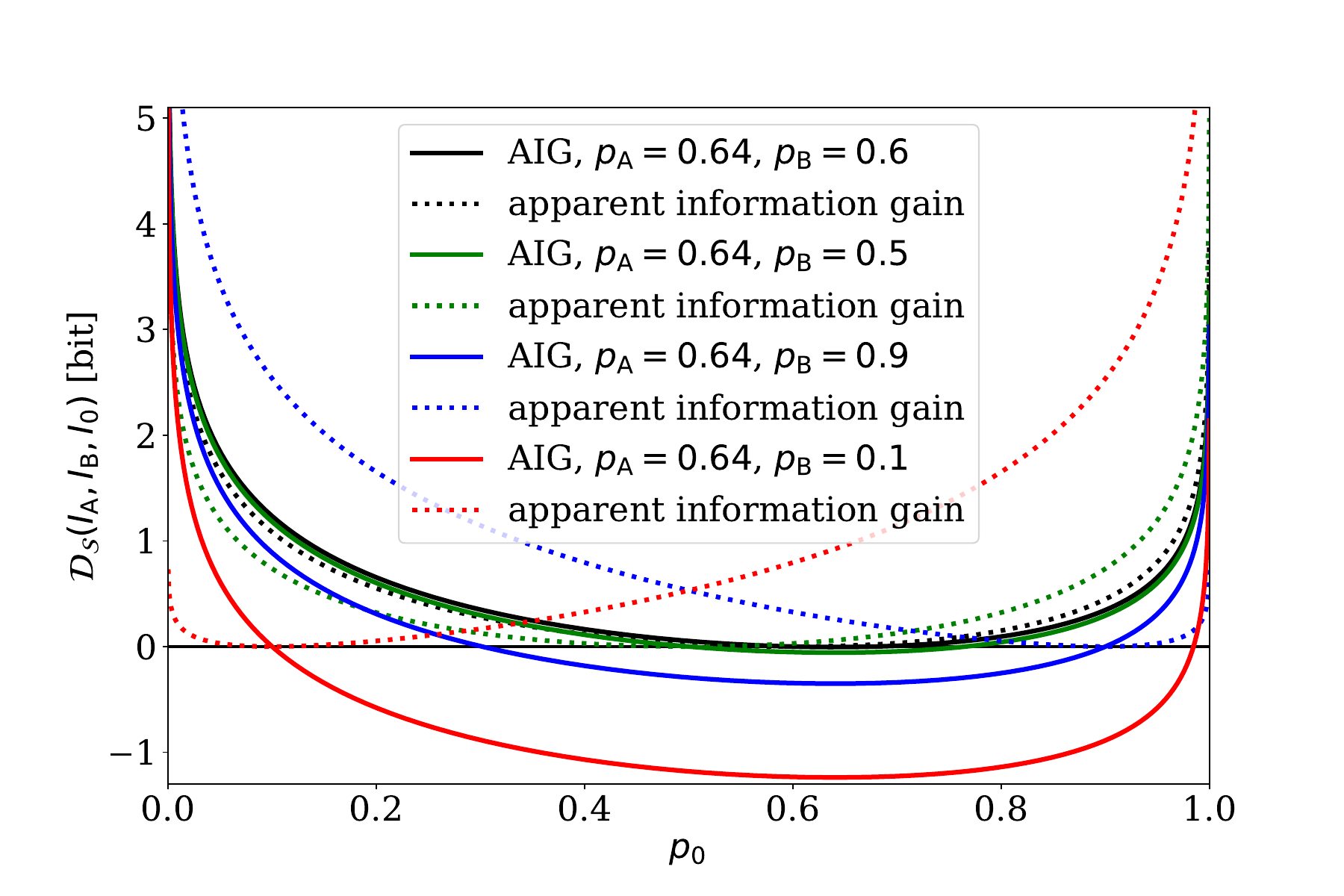}\includegraphics[viewport=15bp 20bp 785bp 510bp,clip,width=0.49\textwidth]{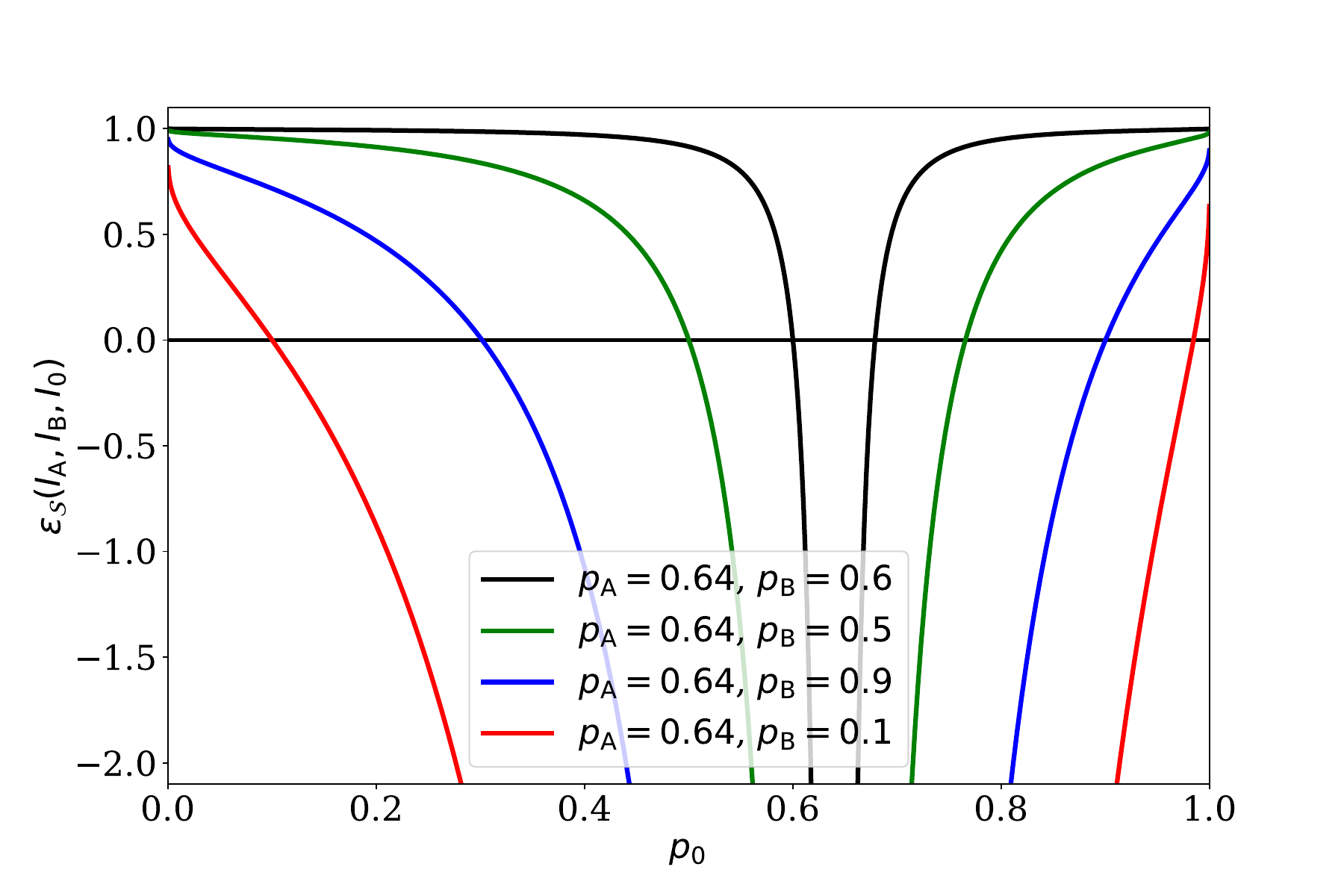}\caption{Knowledge update on a Bernoulli distributed event with occurrence
rate $p_{\text{A}}=0.64$ for various values of the update $p_{\text{B}}$
and as a function of the initial belief $p_{\text{0}}$. Left: AIG
and apparent information gain. Right: Cognitive fidelity. \protect\label{fig:AIG_Bernoulli}}
\end{figure*}
In the following we discuss a number of instructive examples, all
with relevance for practical applications. They rely on having access
to an ideal knowledge state $I_{\text{A}}$ of Alice, as well as the
initial $I_{\text{0}}$ and updated $I_{\text{B}}$ belief of Bob.
In cases when only summaries of the corresponding probability densities
are given, like their mean $m_{\text{X}}:=\langle s\rangle_{(s|I_{\text{X}})}$
and variance $D_{\text{X}}:=\langle(s-m)\,(s-m)^{\text{t}}\rangle_{(s|I_{\text{X}})}$
with $\text{X}\in\{\text{A},\text{B},\text{0}\}$, the Maximum Entropy
Principle \cite{PhysRev.106.620,PhysRev.108.171,jaynes1963information,jaynes1968prior,jaynes03}
permits to construct appropriate PDFs. For example, in case $s\in\mathbb{R}^{n}$
and no other constrain than $I_{\text{X}}=(m_{\text{X}},D_{\text{X}})$
is given, the Maximum Entropy Principle suggests
\begin{eqnarray}
\mathcal{P}(s|I_{\text{X}}) & \!\!\!\!=\!\!\!\! & \mathcal{G}(s-m_{\text{X}},D_{\text{X}})\text{, with}\\
\mathcal{G}(x,X) & \!\!\!\!:=\!\!\!\! & \frac{1}{\sqrt{|2\pi D|}}\exp\left(-\frac{1}{2}x^{\text{t}}X^{-1}x\right)
\end{eqnarray}
as the assumption that adds the least amount of spurious information.
This renders Gaussian PDFs particular useful. For this reason, the
multivariate Gaussian distribution will receive special attention
here.

We start our AIG examples with the simplest possible distribution.

\subsection{Bernoulli distribution}

The simplest possible distribution is the Bernoulli one with only
two outcomes, say $s\in\mathcal{S}=\{0,1\}$. Any knowledge state
$I_{\text{X}}$ on this outcome with $\text{X}\in\{\text{A},\text{B,\ensuremath{0}}\}$
is then given by a single number $p_{\text{X}}=P(s=0|I_{\text{X}})\in[0,1]$,
the by $\text{X}$ expected frequency of outcome $s=0$, as that of
the other possible outcome is $P(s=1|I_{\text{X}})=1-p_{\text{X}}.$
The AIG is then
\begin{equation}
\mathcal{D}_{\mathcal{S}}(I_{\text{A}},I_{\text{B}},I_{\text{0}})=p_{\text{A}}\ln\frac{p_{\text{B}}}{p_{\text{0}}}+(1-p_{\text{A}})\ln\frac{1-p_{\text{B}}}{1-p_{\text{0}}}.\label{eq:AIG-Bernoulli}
\end{equation}

A real world application for this is for example the need of the weather
reporter Alice to decide how important it is to update her audience
about an improved weather forecast. The policy of her station is to
round any communicated probability to 10\%. The announcement, for
which she needs to decide whether to make it, is of a rain probability
of 60\% for the next day. Is this a news worth communicating?

Imagine, the audience is aware of a 50\% rain probability for the
next day, $(p_{\text{0}}=0.5$) and Alice knows that the real updated
probability is $64\%$ ($p_{\text{A}}=0.64$), but she would communicate
$60\%$ ($p_{\text{B}}=0.6$). The AIG of her communication would
then be only $\mathcal{D}_{\mathcal{S}}(I_{\text{A}},I_{\text{B}},I_{\text{0}})=0.052$
bit, letting her likely decide not to make the announcement.

If, however, Alice previously reported a rain risk of only $10\%$
($p_{0}=0.1$), then the AIG of the communication would be $1.23$
bit, as she would correct her audience's belief significantly. This
will probably trigger her to do the announcement. The cognitive fidelity
of her communication act would be $91.5\%$ in the former and $99.6\%$
in the latter case.

Fig.\ \ref{fig:AIG_Bernoulli} shows in the left panel the AIG and
apparent information gain in the scenario with $p_{\text{A}}=0.64$
for $p_{\text{B}}\in\{0.1,0.5,0.6,0.9\}$ as a function of $p_{\text{0}}$.
Two things are worth noting: First, the apparent information gain
is often much larger than the AIG and therefore can be very misleading.
Second, under specific circumstances it is possible that strongly
incorrect statements like a rain probability of 90\%, when it is actually
64\%, can even improve the knowledge of the audience. This happens
in cases in which the initial belief was too much off; here if it
was below 30\% or above 90\%. The cognitive fidelity shown in the
right panel of Fig.\ \ref{fig:AIG_Bernoulli} makes it apparent that
if the initial belief was close to the correct one, an approximate
update is counter productive.
\begin{figure*}
\includegraphics[viewport=45bp 20bp 805bp 510bp,clip,width=0.49\textwidth]{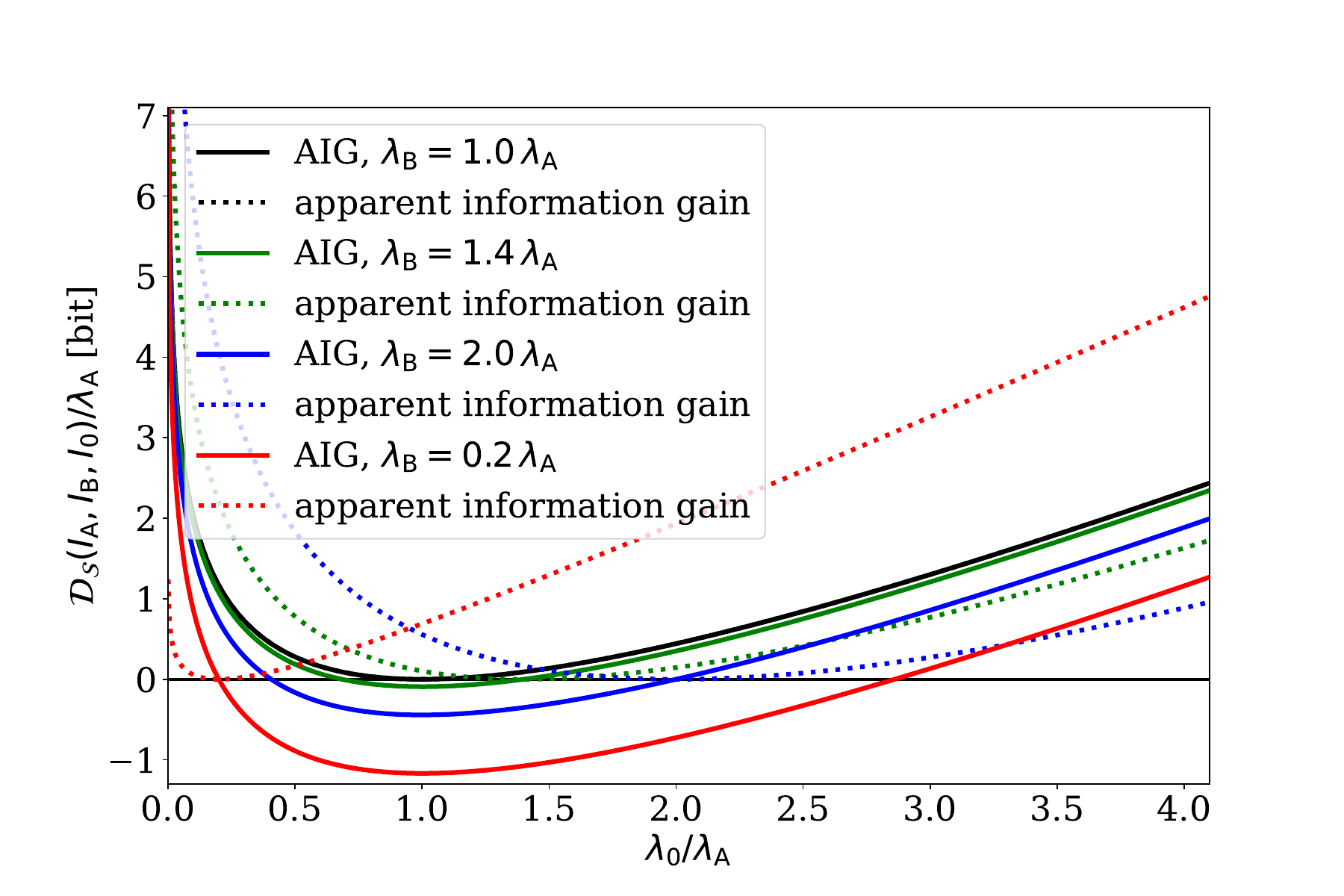}\includegraphics[viewport=15bp 20bp 785bp 510bp,clip,width=0.49\textwidth]{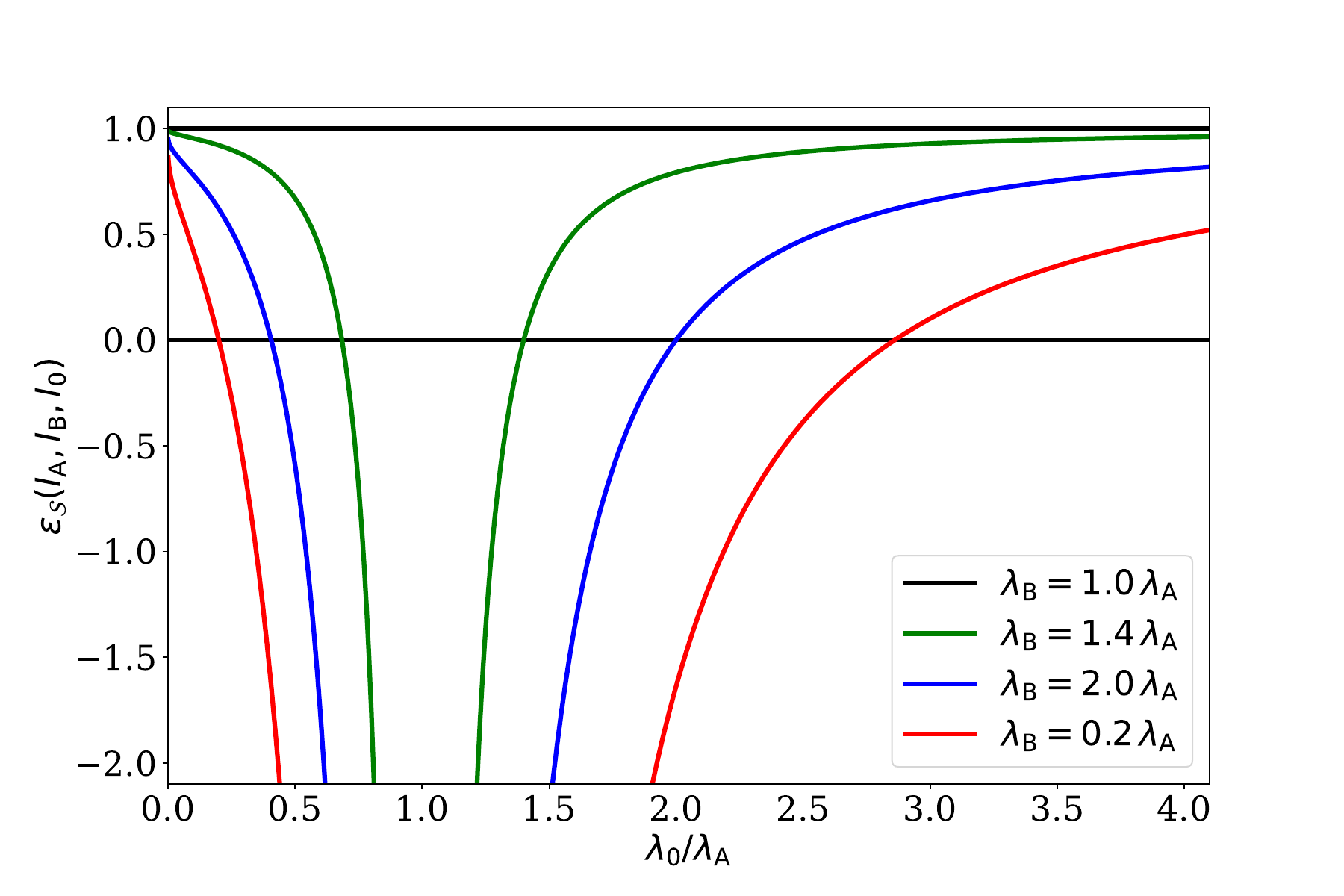}\caption{Knowledge update on a Poisson distributed quantity with expectation
$\lambda_{\text{A}}$ for various values of the update $\lambda_{\text{B}}$
as a function of the initial belief $\lambda_{\text{0}}$. Left: AIG
and apparent information gain. Right: Cognitive fidelity. \protect\label{fig:AIG_Poisson}}
\end{figure*}

\subsection{Binomial distribution}

The binomial distribution plays a central role in the analysis of
repeated Bernoulli medical experiments, where a number of patients
are exposed to some treatment or a placebo and the statistics of the
occurrence of various effects, like curing a disease or side effects,
are taken under the different conditions \parencite{jung2004estimation}.

The $n$-time repeated Bernoulli draws with a success probability
$p_{\text{X}}$ lead to a number $s\in\mathcal{S}=\mathbb{N}_{0}$
of successes. The distribution for $s$ on this outcome under the
knowledge $I_{\text{X}}=p_{x}$ with $\text{X}\in\{\text{A},\text{B,0}\}$
is the binomial distribution
\begin{equation}
P(s|I_{\text{X}})=B(s|p_{\text{X}},n):=\begin{pmatrix}n\\
s
\end{pmatrix}p_{\text{X}}^{s}(1-p_{\text{X}})^{n-s}.
\end{equation}
The AIG of an update from $I_{\text{0}}$ to $I_{\text{B}}$ from
the perspective of $I_{\text{A}}$ is then
\begin{eqnarray}
\mathcal{D}_{\mathcal{S}}(I_{\text{A}},I_{\text{B}},I_{\text{0}}) & \!\!\!\!\!=\!\!\!\!\! & \sum_{s=0}^{n}\begin{pmatrix}n\\
s
\end{pmatrix}p_{A}^{s}(1-p_{A})^{n-s}\times\nonumber \\
 &  & \left[s\,\ln\frac{p_{\text{B}}}{p_{\text{0}}}+(n-s)\,\ln\frac{1-p_{\text{B}}}{1-p_{\text{0}}}\right]\nonumber \\
 & \!\!\!\!\!=\!\!\!\!\! & \sum_{s=0}^{n}p_{A}^{s}(1-p_{A})^{n-s}\times\nonumber \\
 &  & \left[\frac{n!\,\ln\frac{p_{\text{B}}}{p_{\text{0}}}}{(s-1)!\,(n-s)!}+\frac{n!\,\ln\frac{1-p_{\text{B}}}{1-p_{\text{0}}}}{s!\,(n-s-1)!}\right]\nonumber \\
 & \!\!\!\!\!=\!\!\!\!\! & \sum_{s=1}^{n}p_{A}^{s}(1-p_{A})^{n-s}\frac{n!\,\ln\frac{p_{\text{B}}}{p_{\text{0}}}}{(s-1)!\,(n-s)!}\nonumber \\
 &  & +\sum_{s=0}^{n-1}p_{A}^{s}(1-p_{A})^{n-s}\frac{n!\,\ln\frac{1-p_{\text{B}}}{1-p_{\text{0}}}}{s!\,(n-s-1)!}\nonumber \\
 & \!\!\!\!\!=\!\!\!\!\! & \sum_{s=0}^{n-1}p_{A}^{s+1}(1-p_{A})^{n-1-s}\frac{n!\,\ln\frac{p_{\text{B}}}{p_{\text{0}}}}{s!\,(n-s-1)!}\nonumber \\
 &  & +\sum_{s=0}^{n-1}p_{A}^{s}(1-p_{A})^{n-s}\frac{n!\,\ln\frac{1-p_{\text{B}}}{1-p_{\text{0}}}}{s!\,(n-s-1)!}\nonumber \\
 & \!\!\!\!\!=\!\!\!\!\! & n\,\left[p_{\text{A}}\ln\frac{p_{\text{B}}}{p_{\text{0}}}+(1-p_{\text{A}})\ln\frac{1-p_{\text{B}}}{1-p_{\text{0}}}\right]\times\nonumber \\
 &  & \underbrace{\sum_{s=0}^{n-1}B(s|p_{\text{A}},n-1)}_{=1}\nonumber \\
 & \!\!\!\!\!=\!\!\!\!\! & n\,\left[p_{\text{A}}\ln\frac{p_{\text{B}}}{p_{\text{0}}}+(1-p_{\text{A}})\ln\frac{1-p_{\text{B}}}{1-p_{\text{0}}}\right]\!.
\end{eqnarray}
The AIG for a binomial distributed quantity is exactly $n$-times
the AIG in case of the update on a Bernoulli variable, Eq.\ \ref{eq:AIG-Bernoulli},
as the binomial experiment is an $n$-times repeated Bernoulli experiment.

\subsection{Poisson distribution}

The Poisson distribution describes the number of events from a pool
of virtually infinitely many possible events, each with vanishing
probability, but together with a finite total expected number $\lambda_{\text{X}}$,
\begin{equation}
P(s|\lambda_{\text{X}})=\text{Poisson}(s|\lambda_{\text{X}}):=\frac{\lambda_{\text{X}}^{s}}{s!}e^{-\lambda_{\text{X}}}.
\end{equation}
It is used in many contexts, for the number of photons received, of
radioactive decays recorded, of fishes caught in an ocean, or of galaxies
found in a sub-volume of the Universe.

The functional form of the Poisson distribution can be obtained from
the binomial distribution by taking the limit $n\rightarrow\infty$
while enforcing $n\,p_{\text{X}}\rightarrow\lambda_{\text{X}}$. We
identify the knowledge states with the expected Poisson rate, $I_{\text{X}}\equiv\lambda_{X}$.
The Poisson AIG follows from the same limit from the binomial AIG
as
\begin{eqnarray}
\mathcal{D}_{\mathcal{S}}(I_{\text{A}},I_{\text{B}},I_{\text{0}}) & \!\!\!\!\!=\!\!\!\!\! & \lim_{n\rightarrow\infty}\left[\lambda_{\text{A}}\ln\frac{\lambda_{\text{B}}}{\lambda_{\text{0}}}+(n-\lambda_{\text{A}})\ln\frac{1-\nicefrac{\lambda_{\text{B}}}{n}}{1-\nicefrac{\lambda_{\text{0}}}{n}}\right]\nonumber \\
 & \!\!\!\!\!=\!\!\!\!\! & \lambda_{\text{A}}\ln\frac{\lambda_{\text{B}}}{\lambda_{\text{\ensuremath{0}}}}+\!\!\lim_{n\rightarrow\infty}\!\!\left(\!-\frac{\lambda_{\text{B}}}{n}+\frac{\lambda_{\text{0}}}{n}+\mathcal{O}(n^{\!-2})\!\right)n\nonumber \\
 & \!\!\!\!\!=\!\!\!\!\! & \lambda_{\text{A}}\,\ln\frac{\lambda_{\text{B}}}{\lambda_{\text{0}}}-\lambda_{\text{B}}+\lambda_{\text{0}},
\end{eqnarray}
which can be easily verified by a direct calculation:
\begin{eqnarray}
\mathcal{D}_{\mathcal{S}}(I_{\text{A}},I_{\text{B}},I_{\text{0}}) & \!\!\!\!\!=\!\!\!\!\! & \sum_{s=0}^{\infty}\frac{\lambda_{\text{A}}^{s}}{s!}e^{-\lambda_{\text{A}}}\left[s\,\ln\lambda_{\text{B}}-\lambda_{\text{B}}-s\,\ln\lambda_{\text{0}}+\lambda_{\text{0}}\right]\nonumber \\
 & \!\!\!\!\!=\!\!\!\!\! & \langle s\rangle_{(s|\lambda_{\text{A}})}\,\ln\frac{\lambda_{\text{B}}}{\lambda_{\text{0}}}+\langle1\rangle_{(s|\lambda_{\text{A}})}\,\left[-\lambda_{\text{B}}+\lambda_{\text{0}}\right]\nonumber \\
 & \!\!\!\!\!=\!\!\!\!\! & \lambda_{\text{A}}\,\ln\frac{\lambda_{\text{B}}}{\lambda_{\text{0}}}-\lambda_{\text{B}}+\lambda_{\text{0}}.
\end{eqnarray}
Fig.\ \ref{fig:AIG_Poisson} shows the Poisson AIG and cognitive
fidelity for various scenarios. As for the AIG per $\lambda_{\text{A}}$
only the relative rates $x_{\text{X}}:=\lambda_{\text{X}}/\lambda_{\text{A}}$
matter,
\begin{eqnarray}
\frac{\mathcal{D}_{\mathcal{S}}(I_{\text{A}},I_{\text{B}},I_{\text{0}})}{\lambda_{\text{A}}} & \!\!\!\!\!=\!\!\!\!\! & \ln\frac{x_{\text{B}}}{x_{\text{0}}}-x_{\text{B}}+x_{\text{0}},
\end{eqnarray}
 actually this quantity is displayed as function of $x_{\text{0}}$.
Similar to the Bernoulli and binomial cases, an update provides positive
AIG only if the communicated Poisson rate is closer -- in a specific
sense -- to the real one than the original one and the apparent information
gain can be very misleading. The right panel of Fig.\ \ref{fig:AIG_Poisson}
illustrates that communicating the correct value $\lambda_{\text{B}}=\lambda_{\text{A}}$
has of course maximal cognitive fidelity.

\subsection{Beta distribution}

Knowledge on the frequency $f_{0}=\mathcal{P}(s=0|f_{0})\in\mathcal{F}=[0,1]$
of a Bernoulli process can be accumulated from $X$ observed numbers
of events $n_{\text{X,}0}$ and $n_{\text{X,}1}$ for the two cases
$s=0$ and $s=1$, respectively. If the prior on $f_{0}$ is flat,
$\mathcal{P}(f_{0}|f\in\mathcal{F})=1$, the posterior becomes Beta
distributed
\begin{eqnarray}
\mathcal{P}(f_{0}|I_{\text{X}}) & \!\!\!\!\!=\!\!\!\!\! & \text{Beta}(f_{0}|n_{\text{X},0}+1,n_{\text{X,}1}+1)\text{, with}\\
\text{Beta}(x|a,b) & \!\!\!\!\!:=\!\!\!\!\! & \frac{x^{a-1}\,(1-x)^{b-1}}{\text{\ensuremath{\mathcal{B}}(a,b)}}\text{ and}\\
\mathcal{B}(a,b) & \!\!\!\!\!:=\!\!\!\!\! & \int_{0}^{1}\text{d}x\,x^{a-1}\,(1-x)^{b-1}=\frac{\Gamma(a)\Gamma(b)}{\Gamma(a+b)}.
\end{eqnarray}
Here, $I_{\text{X}}=(n_{\text{X,}0},n_{\text{X,}1})$ is the data
vector of the number of observed cases. The parameters need not be
restricted to integer numbers, but can be allowed to be in $(-1,\infty)$.
This freedom is e.g.\ used to represent knowledge states that are
not only based on direct observation.

For example in some agent based computational psychology simulations
\parencite{2022AnP...53400277E}, agents communicate with each other
about their knowledge on some frequencies, not necessarily in an honest
manner. The knowledge states of the agents about a frequency is stored
in form of the parameters of a beta distribution. Agent Alice might
communicate a knowledge state $J_{\text{A}}$ to Bob. $J_{\text{A}}$
might be her actual belief $I_{\text{A}}$, or a lie $I'_{\text{A}}$
she constructed to manipulate Bob. Bob updates then his belief on
that frequency from $I_{\text{0}}$ to $I_{\text{B}}$, taking into
account how much he trusts Alice. The AIG $\mathcal{D}_{\mathcal{F}}(J_{\text{A}},I_{\text{B}},I_{\text{0}})$
then quantifies Bob's knowledge shift from the perspective of Alice
knowledge in case she was honest, or from the perspective of her intention
in case she is manipulative. In both cases, the AIG allows to track
how much Bob's belief is shaped by Alice.

The AIG for such Beta distributions updates is
\begin{eqnarray}
 &  & \!\!\!\!\!\!\!\!\!\!\mathcal{D}_{\mathcal{F}}(I_{\text{A}},I_{\text{B}},I_{\text{0}})\nonumber \\
 & \!\!\!\!\!=\!\!\!\!\! & \left\langle (n_{\text{B,}0}-n_{\text{0,}0})\,\ln f_{0}+(n_{\text{B,}1}-n_{\text{0,}1})\,\ln(1-f_{0})\right\rangle _{(f_{0}|I_{\text{A}})}\nonumber \\
 &  & -\ln\frac{\text{\ensuremath{\mathcal{B}}(\ensuremath{n_{\text{B,}0}}+1,\ensuremath{n_{\text{B,}1}}+1)}}{\text{\ensuremath{\mathcal{B}}(\ensuremath{n_{\text{0,}0}}+1,\ensuremath{n_{\text{0,}1}}+1)}}\nonumber \\
 & \!\!\!\!\!=\!\!\!\!\! & (n_{\text{B,}0}-n_{\text{0,}0})\,\left[\psi(n_{\text{A,}0}+1)-\psi(n_{\text{A,}0}+n_{\text{A,}1}+2)\right]+\nonumber \\
 &  & (n_{\text{B,}1}-n_{\text{0,}1})\,\left[\psi(n_{\text{A,}1}+1)-\psi(n_{\text{A,}0}+n_{\text{A,}1}+2)\right]+\nonumber \\
 &  & \ln\frac{\text{\ensuremath{\mathcal{B}}(\ensuremath{n_{\text{0,}0}}+1,\ensuremath{n_{\text{0,}1}}+1)}}{\text{\ensuremath{\mathcal{B}}(\ensuremath{n_{\text{B,}0}}+1,\ensuremath{n_{\text{B,}1}}+1)}},
\end{eqnarray}
which can be shown using the relation $\langle\ln x\rangle_{\text{Beta}(x|a,b)}=\psi(a)-\psi(a+b)$
with $\psi(a)=\text{d}\ln\Gamma(a)/\text{d}a$ being the digamma function
\parencite{enwiki:1284886665}.

\begin{figure*}
\includegraphics[viewport=40bp 20bp 785bp 510bp,clip,width=0.49\textwidth]{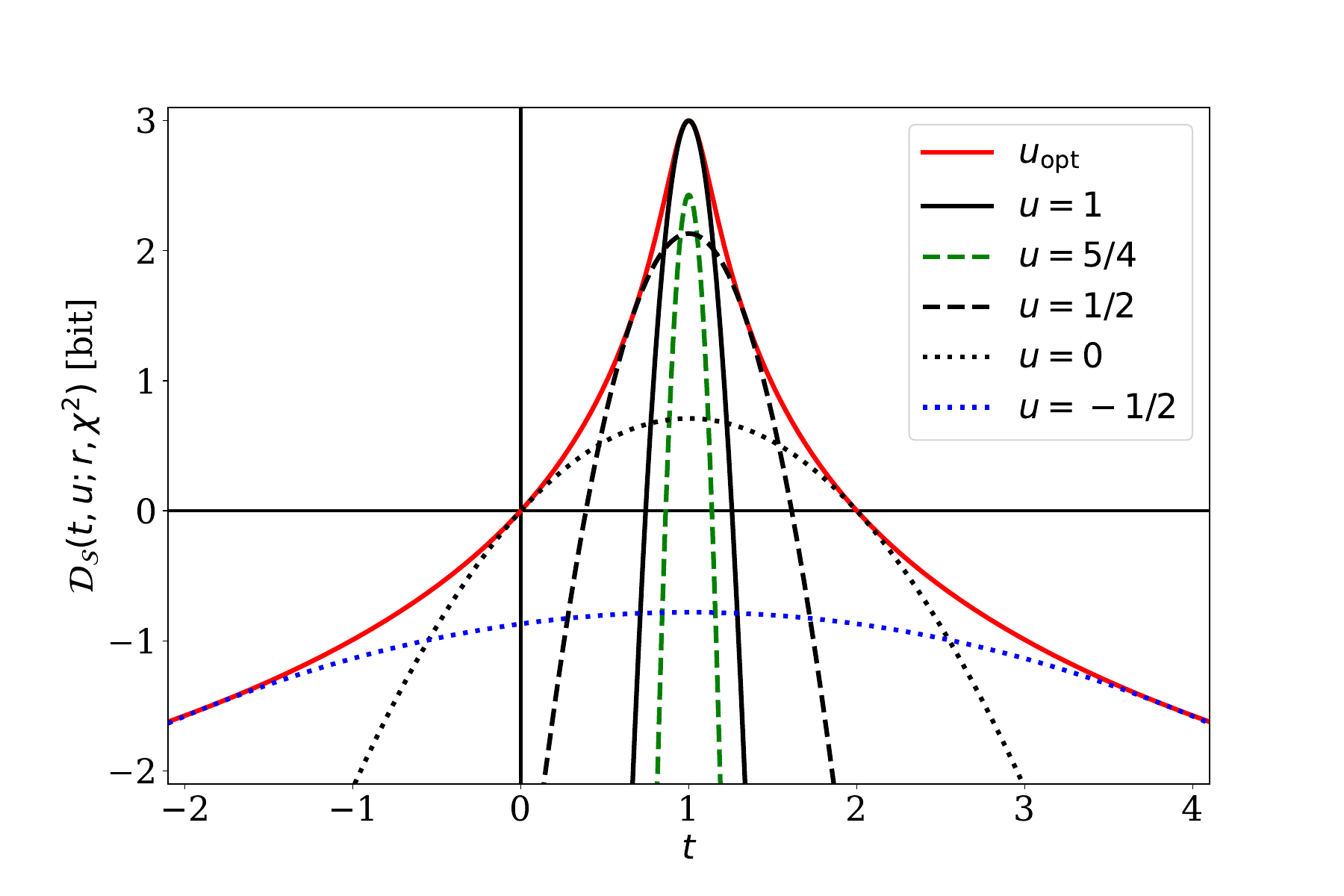}\includegraphics[viewport=35bp 20bp 780bp 510bp,clip,width=0.49\textwidth]{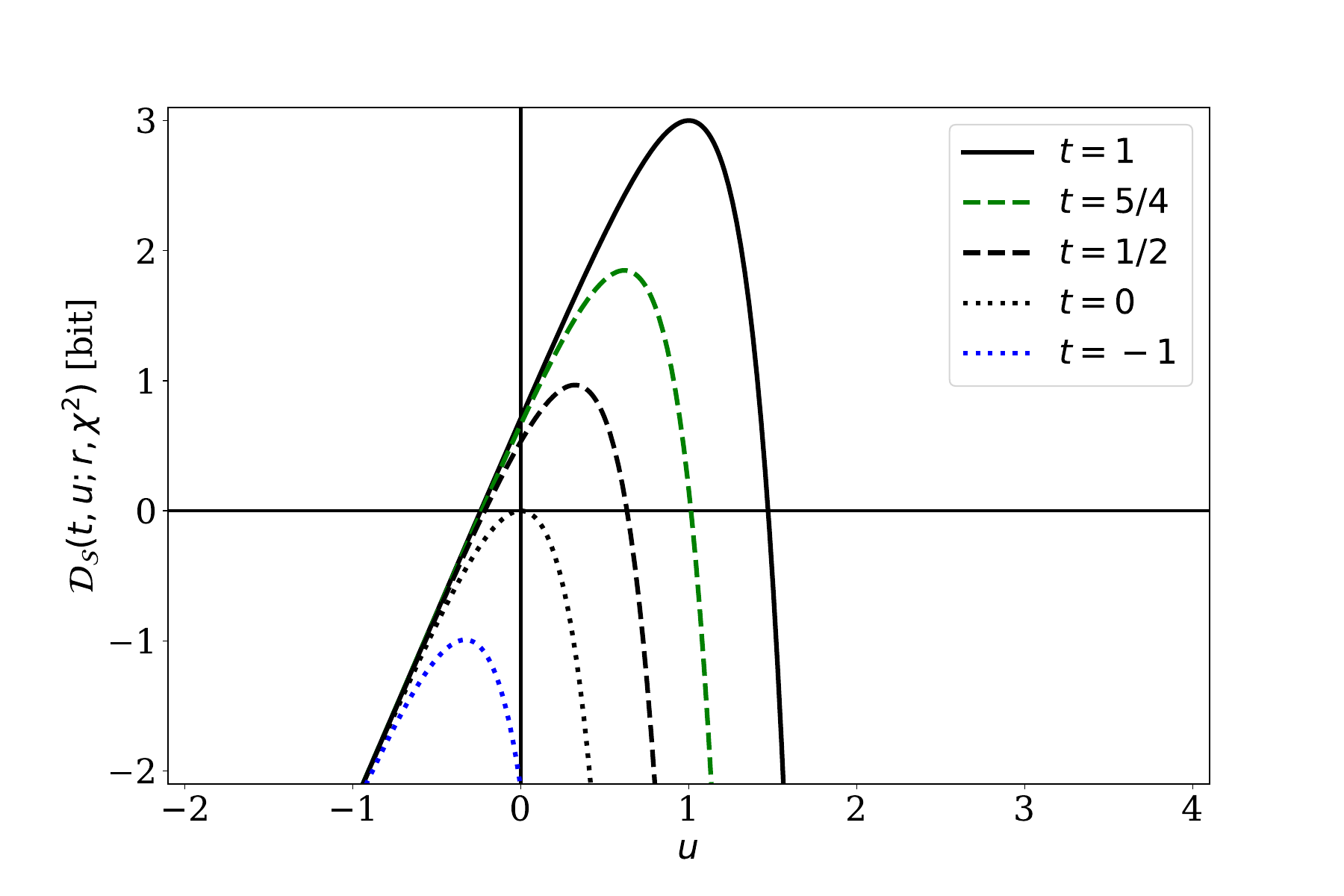}\caption{AIG under incorrect mean (left) and variance (right) of a Gaussian
distribution according to Eq.\ \ref{eq:RIG_IP} as parameterized
by $t$ and $u$, respectively. The other relevant parameters are
$r=2^{-3}$, $\chi^{2}=1-r^{2}$, and $n=1$ and lead to an optimal
information gain of three bit. Also parameter values for $t$ and
$u$ outside the range $[0,1]$ of direct trajectories from $I_{\text{A}}$
to $I_{\text{B}}$ are shown in order to illustrate how quickly incorrect
updates can go into the wrong direction, meaning leading to a negative
AIG. The vanishing initial information gain for $t=u=0$ is marked
by thin black lines. \protect\label{fig:Realized-information-gain-inc-parameter}}
\end{figure*}
how much

\subsection{Multivariate Gaussian probabilities}

A special, but very relevant case is when all involved probabilities
are multivariate Gaussian distribution,
\begin{eqnarray}
\mathcal{P}(s|I_{X}) & = & \mathcal{G}(s-m_{X},D_{X}),\label{eq:Gaussian}
\end{eqnarray}
where $m_{X}$ and $D_{X}$ denote the respective mean and covariance
of the distribution and $X\in\left\{ \text{A},\text{B},\text{0}\right\} $.
In this case the AIG is
\begin{eqnarray}
\mathcal{D_{\mathcal{S}}}(I_{\text{A}},I_{\text{B}},I_{\text{0}}) & \!\!\!\!=\!\!\!\! & \int\text{d}s\,\mathcal{G}(s-m_{\text{A}},D_{\text{A}})\,\ln\frac{\mathcal{G}(s-m_{\text{B}},D_{\text{B}})}{\mathcal{G}(s-m_{\text{0}},D_{\text{0}})}\nonumber \\
 & \!\!\!\!=\!\!\!\! & \int\text{d}s'\,\mathcal{G}(s',D_{\text{A}})\,\ln\!\frac{\mathcal{G}(s'+\Delta_{\text{B}},D_{\text{B}})}{\mathcal{G}(s'+\Delta_{\text{0}},D_{\text{0}})}\nonumber \\
 & \!\!\!\!=\!\!\!\! & \left\langle \ln\frac{\mathcal{G}(s'+\Delta_{\text{B}},D_{\text{B}})}{\mathcal{G}(s'+\Delta_{\text{0}},D_{\text{0}})}\right\rangle _{\mathcal{G}(s',D_{\text{A}})}\nonumber \\
 & \!\!\!\!=\!\!\!\! & \frac{1}{2}\ln\frac{|D_{\text{0}}|}{|D_{\text{B}}|}\nonumber \\
 &  & +\frac{1}{2}\left\langle \left(s'+\Delta_{\text{0}}\right)^{\text{t}}D_{\text{0}}^{-1}\left(s'+\Delta_{\text{0}}\right)\right\rangle _{\mathcal{G}(s',D_{\text{A}})}\nonumber \\
 &  & -\frac{1}{2}\left\langle \left(s'+\Delta_{\text{B}}\right)^{\text{t}}D_{\text{B}}^{-1}\left(s'+\Delta_{\text{B}}\right)\right\rangle _{\mathcal{G}(s',D_{\text{A}})}\nonumber \\
 & \!\!\!\!=\!\!\!\! & \frac{1}{2}\ln\frac{|D_{\text{0}}|}{|D_{\text{B}}|}+\frac{1}{2}\text{Tr}\left[D_{\text{0}}^{-1}\left(D_{\text{A}}+\Delta_{\text{0}}\Delta_{\text{0}}^{\text{t}}\right)\right]\nonumber \\
 &  & -\frac{1}{2}\text{Tr}\left[D_{\text{B}}^{-1}\left(D_{\text{A}}+\Delta_{\text{B}}\Delta_{\text{B}}^{\text{t}}\right)\right]\nonumber \\
 & \!\!\!\!=\!\!\!\! & \mathinner{\color{violet}\frac{1}{2}}{\color{violet}\ln}\mathinner{\color{violet}\frac{|D_{\text{0}}|}{|D_{\text{B}}|}}+\frac{1}{2}\text{Tr}\left[\left({\color{blue}D_{\text{0}}^{-1}}\mathbin{\color{red}-}{\color{red}D_{\text{B}}^{-1}}\right)D_{\text{A}}\right]\nonumber \\
 &  & +\frac{1}{2}\left({\color{teal}\Delta_{\text{0}}^{\text{t}}D_{\text{0}}^{-1}\Delta_{\text{0}}}\mathbin{\color{orange}-}{\color{orange}\Delta_{\text{B}}^{\text{t}}D_{\text{B}}^{-1}\Delta_{\text{B}}}\right)\label{eq:GaussianRIG}\\
 & \!\!\!\!=:\!\!\!\! & \text{I}+\text{II}+\text{III}
\end{eqnarray}
with $s'=s-m_{\text{A}}$, $\Delta_{X}=m_{\text{A}}-m_{X}$, and using
$s'{}^{\text{t}}D^{-1}s'=\text{Tr}\left(D^{-1}s's'{}^{\text{t}}\right)$,
$\left\langle s's'{}^{\text{t}}\right\rangle _{\mathcal{G}(s',D)}=D$,
as well as $\left\langle s'\right\rangle _{\mathcal{G}(s',D)}=0$.
To facilitate the following calculations we call the three main terms
$\text{I}$, $\text{II}$, and $\text{III}$. Furthermore, we colored
some of their sub-terms for easier visual tracking.

The different terms in this expression should be briefly discussed.
The first term can be rewritten as

\begin{equation}
\text{I}:=\mathinner{\color{violet}\frac{1}{2}}{\color{violet}\ln}\mathinner{\color{violet}\frac{|D_{\text{0}}|}{|D_{\text{B}}|}}=\mathinner{\color{violet}\frac{1}{2}}{\color{violet}\text{Tr}\ln}\mathinner{\color{violet}\left(D_{\text{B}}^{-1}D_{\text{0}}\right)}.
\end{equation}
It lets the AIG increase logarithmically with increasing precision
$D_{\text{B}}^{-1}$ of the updated information. In contrast to this,
the $I_{\text{B}}$-dependent part of the second term, $\text{II}:=-\frac{1}{2}\text{Tr}\left[D_{\text{B}}^{-1}D_{\text{A}}\right]+\text{const}$,
decreases linearly with increasing $D_{\text{B}}^{-1}$. Optimizing
the sum of these two terms for $D_{\text{B}}^{-1}$ yields 
\begin{eqnarray}
0 & = & \frac{\partial\text{I}+\text{II}}{\partial D_{\text{B}}^{-1}}=\frac{1}{2}\left[{\color{violet}D_{\text{B}}}\mathbin{\color{red}-}{\color{red}D_{\text{A}}}\right]
\end{eqnarray}
from which $D_{\text{B}}=D_{\text{A}}$ would follow for the maximal
AIG, but only if there is no relevant contribution from the third
term. The $I_{\text{B}}$-dependent part of it reads $\text{III}:=\text{const}-\frac{1}{2}\Delta_{\text{B}}^{\text{t}}D_{\text{B}}^{-1}\Delta_{\text{B}}$
and is maximal for $\Delta_{\text{B}}=m_{\text{A}}-m_{\text{B}}=0$,
which is the case for $m_{\text{B}}=m_{\text{A}}$.

To summarize, optimal AIG requires -- unsurprisingly -- that mean
and variance of the ideal Gaussian posterior are matched. An error
in the covariance, $D_{\text{B}}\neq D_{\text{A}}$, still lets the
correct mean to be preferred, as the term $\text{III}$ wants $\Delta_{\text{B}}=m_{\text{A}}-m_{\text{B}}=0$
for any positive definite precision matrix $D_{\text{B}}^{-1}$. The
opposite is, however, not correct. An offset between ideal and achieved
posterior mean, $\Delta_{\text{B}}\neq0$, asks for a reduced approximate
precision matrix, in order to accommodate the error made in an increased
uncertainty budget. In case of such an offset, the optimal approximate
uncertainty covariance follows from 
\begin{eqnarray}
0 & = & \frac{\partial\mathcal{D_{\mathcal{S}}}(I_{\text{A}},I_{\text{B}},I_{\text{0}})}{\partial D_{\text{B}}^{-1}}\nonumber \\
 & = & \frac{1}{2}\left[{\color{violet}D_{\text{B}}}\mathbin{\color{red}-}{\color{red}D_{\text{A}}}\mathbin{\color{orange}-}{\color{orange}\Delta_{\text{B}}\Delta_{\text{B}}^{\text{t}}}\right]
\end{eqnarray}
to be ${\color{violet}D_{\text{B}}}={\color{red}D_{\text{A}}}\mathbin{\color{orange}+}{\color{orange}\Delta_{\text{B}}\Delta_{\text{B}}^{\text{t}}}$,
the ideal covariance plus a correction term for the approximation
error in the mean. Typically, this will not be known, but might be
replaced with an appropriate expectation value $\left\langle \Delta_{\text{B}}\Delta_{\text{B}}^{\text{t}}\right\rangle _{\mathcal{P}(\Delta_{\text{B}})}$.

\subsection{Incorrect Gaussian parameters}

In order to see how sensitive the AIG is to incorrect parameters of
$I_{\text{B}}$ in case of Gaussian distributions, we assume 
\begin{eqnarray}
m_{\text{B}} & = & t\,m_{\text{A}}+(1-t)\,m_{\text{0}}\\
D_{\text{B}} & = & D_{\text{A}}^{u/2}\,D_{\text{0}}^{1-u}D_{\text{A}}^{u/2}
\end{eqnarray}
for some $t,u\in\mathbb{R}$. These parameterize paths of $I_{\text{B}}$
from $I_{\text{0}}$ ($t=u=0$) to $I_{\text{A}}$ ($t=u=1$). This
way, we can examine the AIG as a function of the approximation errors
in $m_{\text{B}}$ and $D_{\text{B}}$ as parameterized by $t$ and
$u$, respectively:

\begin{eqnarray}
\mathcal{D_{\mathcal{S}}}(I_{\text{A}},I_{\text{B}},I_{\text{0}}) & \!\!\!\!=\!\!\!\! & \frac{1}{2}\text{Tr}\left[\mathop{\color{violet}\ln}\mathinner{\color{violet}\left(D_{\text{A}}^{-u}\,D_{\text{0}}^{u}\right)}\mathbin{\color{red}-}{\color{red}D_{\text{A}}^{1-u}\,D_{\text{0}}^{u-1}}\right]\nonumber \\
 &  & +\frac{1}{2}\text{Tr}\left[\left({\color{blue}D_{\text{A}}}\mathbin{\color{teal}+}{\color{teal}\Delta_{\text{0}}\Delta_{\text{0}}^{\text{t}}}\right)D_{\text{0}}^{-1}\right]\\
 &  & \mathbin{\color{orange}-}{\color{orange}\frac{1}{2}(1-t)^{2}\Delta_{\text{0}}^{\text{t}}D_{\text{A}}^{-u/2}\,D_{\text{0}}^{u-1}D_{\text{A}}^{-u/2}\Delta_{\text{0}}},\nonumber \\
\text{since }\Delta_{\text{B}} & \!\!\!\!=\!\!\!\! & (1-t)\,(m_{\text{A}}-\,m_{\text{0}})=(1-t)\,\Delta_{\text{0}}^{\text{t}}
\end{eqnarray}
For illustrative purposes, let us further assume that $D_{\text{A}}=r^{2}\,D_{\text{0}}$
with $r\in(0,1]$ describing a linear uncertainty reduction of the
ideal update in every of the $n$ dimension of $s$. If we further
define $\chi^{2}:=\Delta_{\text{0}}^{\text{t}}D_{\text{0}}^{-1}\Delta_{\text{0}}/n\in\mathbb{R^{+}}$,
the squared shift of the ideal mean in terms of prior one-sigma levels,
basically a $\chi^{2}$-value per degree of freedom of the update,
we find that

\begin{eqnarray}
\mathcal{D_{\mathcal{S}}}(t,u;r,\chi^{2})\!\!\!\!\! & = & \!\!\!\!\!\frac{n}{2}\left[\mathbin{\color{violet}-}{\color{violet}2\,u\ln r}\mathbin{\color{red}-}{\color{red}r^{2-2u}}\mathbin{\color{blue}+}{\color{blue}r^{2}}\right.\nonumber \\
 &  & \left.+\left({\color{teal}1}\mathbin{\color{orange}-}{\color{orange}(1-t)^{2}r^{-2u}}\right)\chi^{2}\right].\label{eq:RIG_IP}
\end{eqnarray}

\begin{figure}[t]
\includegraphics[viewport=20bp 20bp 800bp 510bp,clip,width=0.49\textwidth]{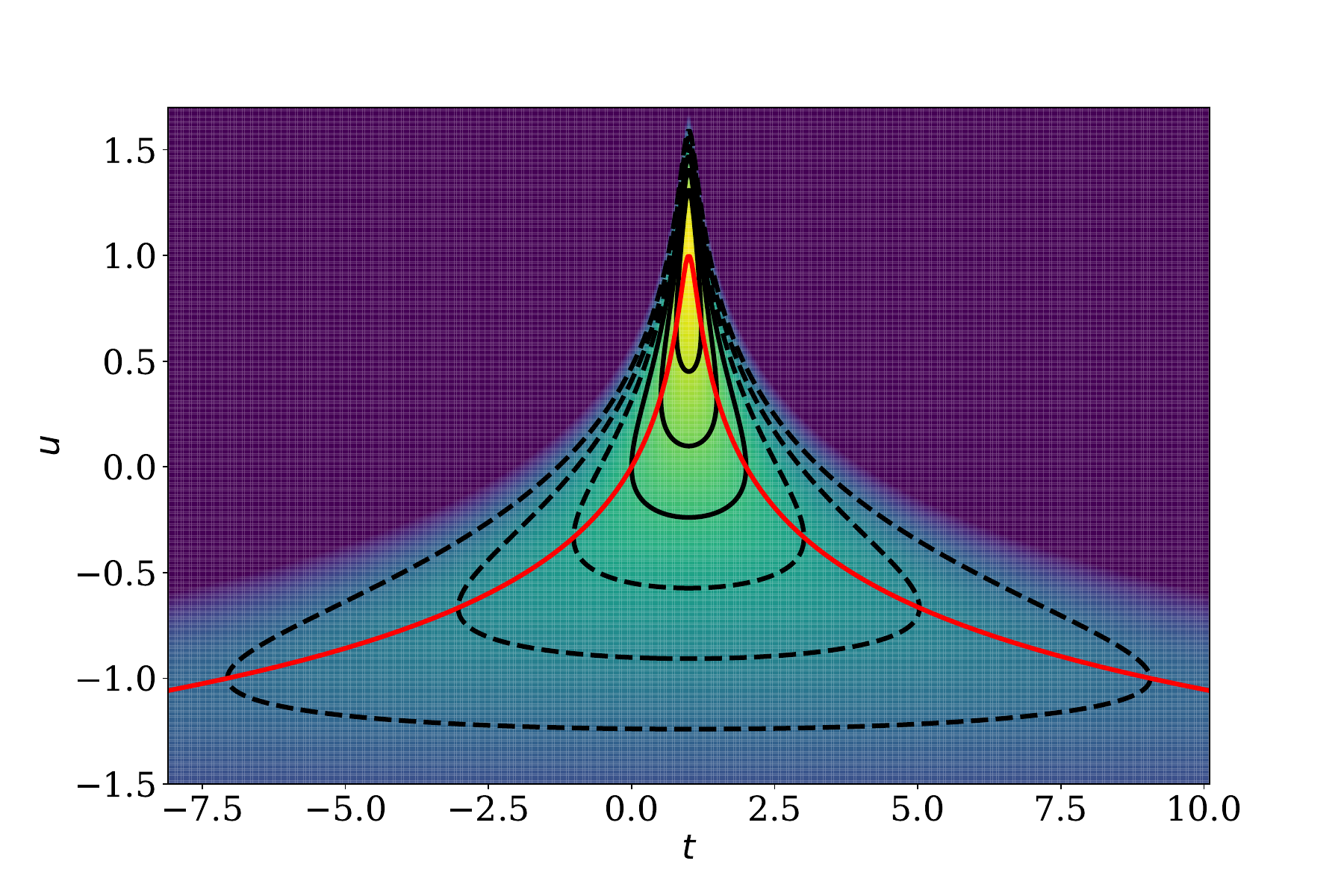}\caption{AIG under incorrect mean and variance for a Gaussian distribution
according to Eq.\ \ref{eq:RIG_IP} as parameterized by $t$ and $u$,
respectively. The other relevant parameters are like in Fig.\ \ref{fig:Realized-information-gain-inc-parameter},
$r=2^{-3}$, $\chi^{2}=1-r^{2}$, and $n=1$. The color varies in
the interval $[-6,3]\text{ bit}$. Contours are shown for $\{-3,\ldots\,2\}\text{ bit}$
and are dashed in the negative range. The red line marks $u_{\text{opt}}(t)$
according to Eq.\ \ref{eq:uopt}. Its maximum is at the global maximum
of the AIG of three bit. \protect\label{fig:Realized-information-gain-2d}}
\end{figure}

Some instructive views on this function are given by Figs.\ \ref{fig:Realized-information-gain-inc-parameter}
and \ref{fig:Realized-information-gain-2d}. The achieved gains at
the initial $(0,0)$ and the end point $(1,1)$ of the trajectory
of $(t,u)$ are

\begin{eqnarray}
\mathcal{D_{\mathcal{S}}}(0,0;r,\chi^{2})\!\!\!\!\! & = & \!\!\!\!\!0\\
\mathcal{D_{\mathcal{S}}}(1,1;r,\chi^{2})\!\!\!\!\! & = & \!\!\!\!\!\frac{n}{2}\left[{\color{olive}\chi^{2}}{\color{violet}\mathbin{\color{red}-}{\color{red}1}\mathbin{\color{blue}+}{\color{blue}r^{2}}-2\,\ln r}\right]\\
\!\!\!\!\! & \approx & \!\!\!\!\!n\ln r^{-1},
\end{eqnarray}
where for the approximation we used that the shift per degree of freedom
is typically\footnote{The ground truth would have an expected reduced $\chi^{2}$ distance
of one from the initial or prior mean. The posterior, however, is
always a bit shifted towards the mean of the prior, explaining the
``$-r^{2}$'' term.} $\chi^{2}\approx1-r^{2}$. Thus, in a typical ($\chi^{2}\approx1-r^{2})$
and ideal ($t=u=1$) measurement the AIG is $m\times n$ bits, where
$m=\log_{2}r^{-1}$ counts how often the uncertainty is halved for
each of the $n$ degrees of freedom. Any difference of $\chi^{2}$
from $1-r^{2}$ imprints on the actual achievable total information
gain.

The approximate uncertainty $D_{\text{B}}=r^{2u}D_{\text{0}}$ becomes
optimal if 
\begin{equation}
0=\frac{\partial\mathcal{D_{\mathcal{S}}}}{\partial u}=n\left(\mathbin{\color{violet}-}{\color{violet}1}+{\color{red}r^{2-2u}}+\mathopen{\color{orange}(}{\color{orange}1-t)^{2}r^{-2u}\chi^{2}}\right)\ln r,
\end{equation}
which for any $r\neq1$ is
\begin{eqnarray}
{\color{violet}u}_{\text{opt}}(t) & = & \frac{\ln\left({\color{red}r^{2}}+\mathopen{\color{orange}(}{\color{orange}1-t)^{2}\chi^{2}}\right)}{\mathop{\color{violet}\ln}{\color{violet}r^{2}}}\text{, implying}\label{eq:uopt}\\
D_{\text{B}}^{\text{opt}}(t) & = & r^{2u}D_{\text{0}}=\left({\color{red}r^{2}}+\mathopen{\color{orange}(}{\color{orange}1-t)^{2}\chi^{2}}\right)D_{\text{0}}\nonumber \\
 & = & D_{\text{A}}+\mathopen{\color{orange}(}{\color{orange}1-t)^{2}\chi^{2}}D_{\text{0}}.
\end{eqnarray}
This means that in case of an incorrect posterior mean, ideally the
uncertainty covariance is enlarged to keep as much information as
possible. $u_{\text{opt}}(t)$ is shown in Fig.\ \ref{fig:Realized-information-gain-2d}
as a red line and the corresponding optimal information gain, $\mathcal{D_{\mathcal{S}}}(t,u_{\text{opt}}(t);r,\chi^{2})$,
is displayed in the left panel of Fig.\ \ref{fig:Realized-information-gain-inc-parameter}
by a red line as well.

In case of an imperfect chosen posterior variance, as parameterized
by $u$, the optimal mean is still the correct one, as expressed by
$t=1$. Thus, the corresponding optimal information gain is $\mathcal{D_{\mathcal{S}}}(1,u;r,\chi^{2})$,
which is shown in the right panel Fig.\ \ref{fig:Realized-information-gain-inc-parameter}.

\begin{figure}
\includegraphics[viewport=45bp 20bp 805bp 510bp,clip,width=0.49\textwidth]{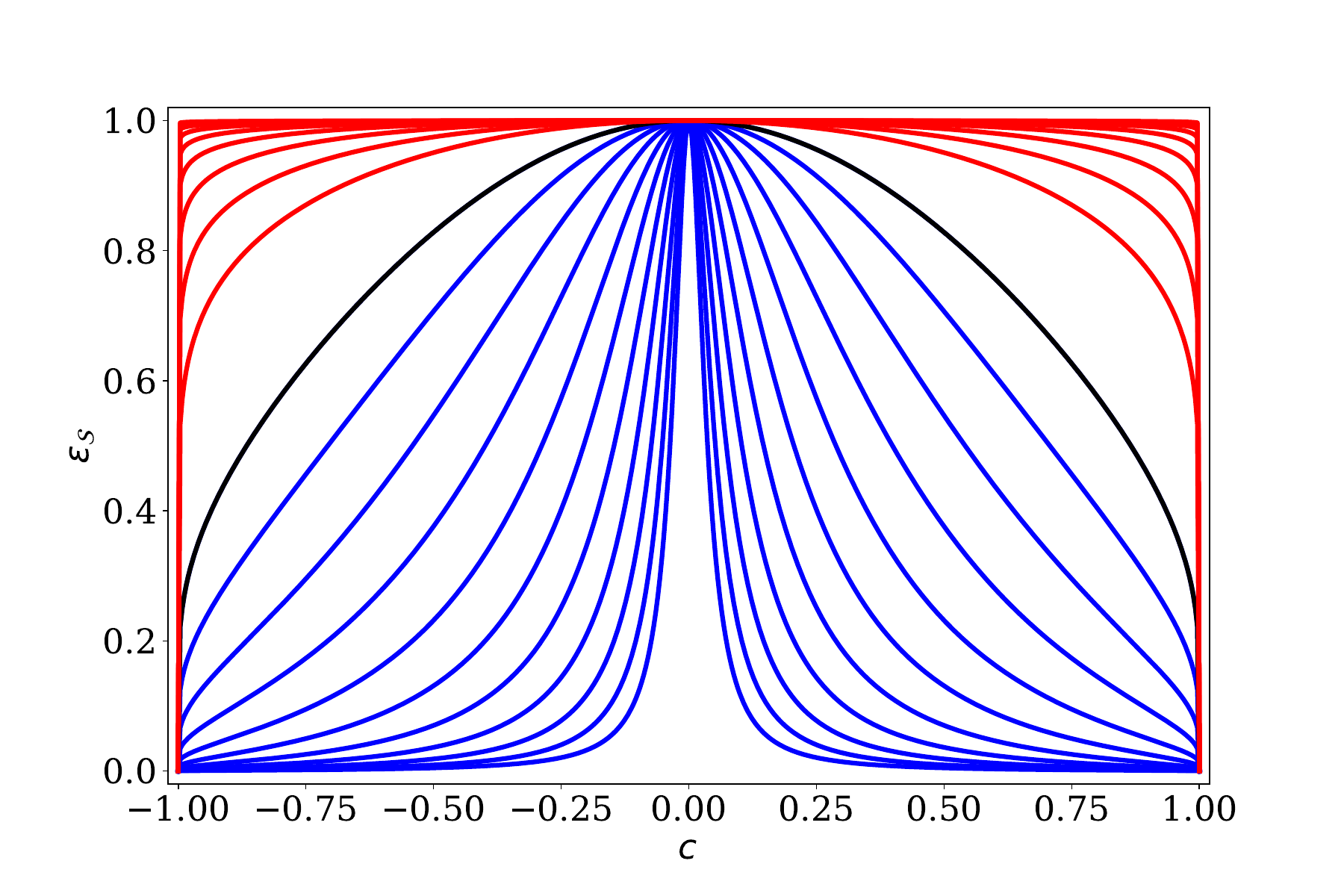}\caption{Cognitive fidelity in case of the mean field approximation scenario
discussed in Sect.\ \ref{subsec:Mean-Field-Approximation}. $\varepsilon_{\mathcal{S}}(I_{\text{A}},I_{\text{B}},I_{\text{0}})$
as given by Eq.\ \ref{eq:MFCogEff} is displayed as a function of
the neglected correlation coefficient $c$ for various amounts of
apparent information gain and AIG $\mathcal{D_{S}}(I_{\text{B}},I_{\text{0}})=\mathcal{D_{S}}(I_{\text{A}},I_{\text{B}},I_{\text{0}})=2^{i}\text{ bit}$
with $i\in\{-10,\ldots,10\}$. The curves for negative values of $i$
are plotted in blue, the ones for positive values in red, and the
one for $i=0$ in black. \protect\label{fig:MFCogEff}}
\end{figure}

\subsection{Mean Field Approximation\protect\label{subsec:Mean-Field-Approximation}}

An approximation often used in inference is the mean field approximation,
in which posterior correlations of parameters are ignored. This correspond
in case of Gaussian distributions to $m_{\text{B}}=m_{\text{A}}$,
but $D_{\text{B}}=\widehat{\widehat{D_{\text{A}}}}=\left[\left(D_{\text{A}}\right)_{ii}\delta_{ij}\right]_{ij}$.
Here, a hat on a matrix turns it into a vector with the matrix diagonal
elements as its components, and a hat on a vector turns the latter
back into a diagonal matrix, with the vector providing the diagonal
elements. Thus, a double hat sets all non-diagonal elements of a matrix
to zero.

According to Eq.\ \ref{eq:GaussianRIG}, the AIG is then
\begin{eqnarray}
\mathcal{D_{\mathcal{S}}}(I_{\text{A}},I_{\text{B}},I_{\text{0}}) & \!\!\!\!=\!\!\!\! & \mathinner{\color{violet}\frac{1}{2}}{\color{violet}\ln}\mathinner{\color{violet}\frac{|D_{\text{0}}|}{|\widehat{\widehat{D_{\text{A}}}}|}}+\frac{1}{2}\text{Tr}\left[\left({\color{blue}D_{\text{0}}^{-1}}-{\color{red}\widehat{\widehat{D_{\text{A}}}}^{-1}}\right)D_{\text{A}}\right]\nonumber \\
 &  & \mathbin{\color{teal}+}{\color{teal}\frac{1}{2}\Delta_{\text{0}}^{\text{t}}D_{\text{0}}^{-1}\Delta_{\text{0}}}.
\end{eqnarray}
The ideal information gain is
\begin{eqnarray}
\mathcal{D_{\mathcal{S}}}(I_{\text{A}},I_{\text{0}}) & \!\!\!\!=\!\!\!\! & \mathinner{\color{violet}\frac{1}{2}}{\color{violet}\ln}\mathinner{\color{violet}\frac{|D_{\text{0}}|}{|D_{\text{A}}|}}\mathbin{\color{blue}+}{\color{blue}\frac{1}{2}\text{Tr}}\mathinner{\color{blue}\left[D_{\text{0}}^{-1}D_{\text{A}}\right]}\mathbin{\color{red}-}\mathinner{\color{red}\frac{n}{2}}\nonumber \\
 &  & \mathbin{\color{teal}+}{\color{teal}\frac{1}{2}\Delta_{\text{0}}^{\text{t}}D_{\text{0}}^{-1}\Delta_{\text{0}}}.
\end{eqnarray}
and the remaining information gain is 
\begin{eqnarray}
\mathcal{D_{\mathcal{S}}}(I_{\text{A}},I_{\text{B}}) & \!\!\!\!=\!\!\!\! & \frac{1}{2}\left(\mathop{\color{violet}\ln}\mathinner{\color{violet}\frac{|\widehat{\widehat{D_{\text{A}}}}|}{|D_{\text{A}}|}}\mathbin{\color{blue}+}{\color{blue}\text{Tr}}\mathinner{\color{blue}\left[\widehat{\widehat{D_{\text{A}}}}^{-1}D_{\text{A}}\right]}\mathbin{\color{red}-}{\color{red}n}\right).
\end{eqnarray}
This can be verified by a direct calculation.\footnote{The calculation is
\begin{eqnarray*}
\mathcal{D_{\mathcal{S}}}(I_{\text{A}},I_{\text{B}}) & \!\!\!\!=\!\!\!\! & \left\langle \ln\frac{\mathcal{G}(s',D_{\text{A}})}{\mathcal{G}(s',\widehat{\widehat{D_{\text{A}}}})}\right\rangle _{\mathcal{G}(s',D_{\text{A}})}\\
 & \!\!\!\!=\!\!\!\! & \frac{1}{2}\left\langle \ln\frac{|\widehat{\widehat{D_{\text{A}}}}|}{|D_{\text{A}}|}+\left[s'{}^{\dagger}\widehat{\widehat{D_{\text{A}}}}^{-1}s'-s'{}^{\dagger}D_{\text{A}}^{-1}s'\right]\right\rangle _{\mathcal{G}(s',D_{\text{A}})}\\
 & \!\!\!\!=\!\!\!\! & \frac{1}{2}\ln\frac{|\widehat{\widehat{D_{\text{A}}}}|}{|D_{\text{A}}|}+\frac{1}{2}\text{Tr}\left[\left(\widehat{\widehat{D_{\text{A}}}}^{-1}-D_{\text{A}}^{-1}\right)D_{\text{A}}\right]\\
 & \!\!\!\!=\!\!\!\! & \frac{1}{2}\left(\ln\frac{|\widehat{\widehat{D_{\text{A}}}}|}{|D_{\text{A}}|}+\text{Tr}\left[\widehat{\widehat{D_{\text{A}}}}^{-1}D_{\text{A}}\right]-n\right).
\end{eqnarray*}
}

To have an illustrative example in $n=2$ dimensions, let us assume
$D_{\text{0}}=\mathbb{1}_{2}\in\mathbb{R}^{2\times2}$, 
\begin{equation}
D_{\text{A}}=\begin{pmatrix}1 & c\\
c & 1
\end{pmatrix}\sigma_{\text{A}}^{2}
\end{equation}
with $c,\sigma_{\text{A}}^{2}<1$, and therefore $D_{\text{B}}=\widehat{\widehat{D_{\text{A}}}}=\mathbb{1}_{2}\sigma_{\text{A}}^{2}$.

The AIG is then 
\begin{eqnarray}
\mathcal{D_{\mathcal{S}}}(I_{\text{A}},I_{\text{B}},I_{\text{0}}) & = & \mathcal{D_{\mathcal{S}}}(I_{\text{A}},I_{\text{0}})-\mathcal{D_{\mathcal{S}}}(I_{\text{A}},I_{\text{B}})\nonumber \\
 & = & \mathbin{\color{violet}-}{\color{violet}\ln\sigma_{\text{A}}^{2}}\mathbin{\color{blue}+}{\color{blue}\sigma_{\text{A}}^{2}}\mathbin{\color{red}-}{\color{red}1}\mathbin{\color{teal}+}{\color{teal}\frac{1}{2}\Delta_{\text{0}}^{\text{t}}\Delta_{\text{0}}},
\end{eqnarray}
where the ideal information gain is 
\begin{eqnarray}
\mathcal{D_{\mathcal{S}}}(I_{\text{A}},I_{\text{0}}) & = & \mathbin{\color{violet}-}{\color{violet}\ln}\mathinner{\color{violet}\left[\sigma_{\text{A}}^{2}\sqrt{1-c^{2}}\right]}\mathbin{\color{blue}+}{\color{blue}\sigma_{\text{A}}^{2}}\mathbin{\color{red}-}{\color{red}1}\mathbin{\color{teal}+}{\color{teal}\frac{1}{2}\Delta_{\text{0}}^{\text{t}}\Delta_{\text{0}}},\nonumber \\
\label{eq:idealIGforcorrelations}
\end{eqnarray}
and the remaining information gain is

\begin{eqnarray}
\mathcal{D_{\mathcal{S}}}(I_{\text{A}},I_{\text{B}}) & = & -\ln\sqrt{1-c^{2}}.
\end{eqnarray}
The apparent information gain is 
\begin{eqnarray}
\mathcal{D}_{\mathcal{S}}(I_{\text{B}},I_{\text{0}}) & \!\!\!\!=\!\!\!\! & \mathbin{\color{violet}-}{\color{violet}\ln}\mathinner{\color{violet}\left[\sigma_{\text{A}}^{2}\right]}\mathbin{\color{blue}+}{\color{blue}\sigma_{\text{A}}^{2}}\mathbin{\color{red}-}{\color{red}1}\mathbin{\color{teal}+}{\color{teal}\frac{1}{2}\Delta_{\text{0}}^{\text{t}}\Delta_{\text{0}}}
\end{eqnarray}
as can be obtained by setting $c=0$ in the formula of the ideal gain,
Eq.\ \ref{eq:idealIGforcorrelations}.

We have here one of the rare cases in which $I_{\text{0}}$, $I_{\text{B}}$,
and $I_{\text{A}}$ are aligned, in the sense that according to their
``distances'' $I_{\text{B}}$ seems to lie directly on the line
from $I_{\text{0}}$ to $I_{\text{A}}$:
\begin{equation}
\mathcal{D_{\mathcal{S}}}(I_{\text{A}},I_{\text{0}})=\mathcal{D_{\mathcal{S}}}(I_{\text{A}},I_{\text{B}})+\mathcal{D}_{\mathcal{S}}(I_{\text{B}},I_{\text{0}})
\end{equation}

Thus, the apparent information gain is actually the AIG in this case,
\begin{eqnarray}
\mathcal{D}_{s}(I_{\text{A}},I_{\text{B}},I_{\text{0}}) & \!\!\!\!=\!\!\!\! & \mathcal{D_{\mathcal{S}}}(I_{\text{A}},I_{\text{0}})-\mathcal{D_{\mathcal{S}}}(I_{\text{A}},I_{\text{B}})\nonumber \\
 & \!\!\!\!=\!\!\!\! & \mathcal{D}_{\mathcal{S}}(I_{\text{B}},I_{\text{0}}).
\end{eqnarray}

In terms of this, the cognitive fidelity is therefore

\begin{eqnarray}
\mathcal{\varepsilon}_{s}(I_{\text{A}},I_{\text{B}},I_{\text{0}}) & \!\!\!\!=\!\!\!\! & \frac{\mathcal{D}_{\mathcal{S}}(I_{\text{B}},I_{\text{0}})}{\mathcal{D_{\mathcal{S}}}(I_{\text{A}},I_{\text{B}})+\mathcal{D}_{\mathcal{S}}(I_{\text{B}},I_{\text{0}})}\nonumber \\
 & \!\!\!\!=\!\!\!\! & \left[1+\frac{\mathcal{D}_{\mathcal{S}}(I_{\text{A}},I_{\text{B}})}{\mathcal{D}_{\mathcal{S}}(I_{\text{B}},I_{\text{0}})}\right]^{-1}\nonumber \\
 & \!\!\!\!=\!\!\!\! & \left[1+\frac{\mathbin{\color{violet}-}{\color{violet}\ln\sqrt{1-c^{2}}}}{\mathcal{D}_{\mathcal{S}}(I_{\text{B}},I_{\text{0}})}\right]^{-1}.\label{eq:MFCogEff}
\end{eqnarray}

This is displayed in Fig.\ \ref{fig:MFCogEff} for various values
of $\mathcal{D}_{\mathcal{S}}(I_{\text{B}},I_{\text{0}})$. As is
apparent in this figure, the ignorance of correlations matters mostly
when the AIG is low, as only then learning about the correlation provides
significant further information. This might helps to understand in
which situations the mean field approximation is appropriate and in
which not. Note that in most applications of the mean field approximation
the mean of $s$ is also affected by the approximation, and not only
the uncertainty covariance as we assumed here. The cognitive fidelity
will therefore be usually lower than given by Eq.\ \ref{eq:MFCogEff}
and displayed in Fig.\ \ref{fig:MFCogEff}.

\subsection{Incomplete data usage\protect\label{subsec:Incomplete-data-usage}}

\begin{figure*}
\includegraphics[viewport=40bp 20bp 800bp 510bp,clip,width=0.49\textwidth]{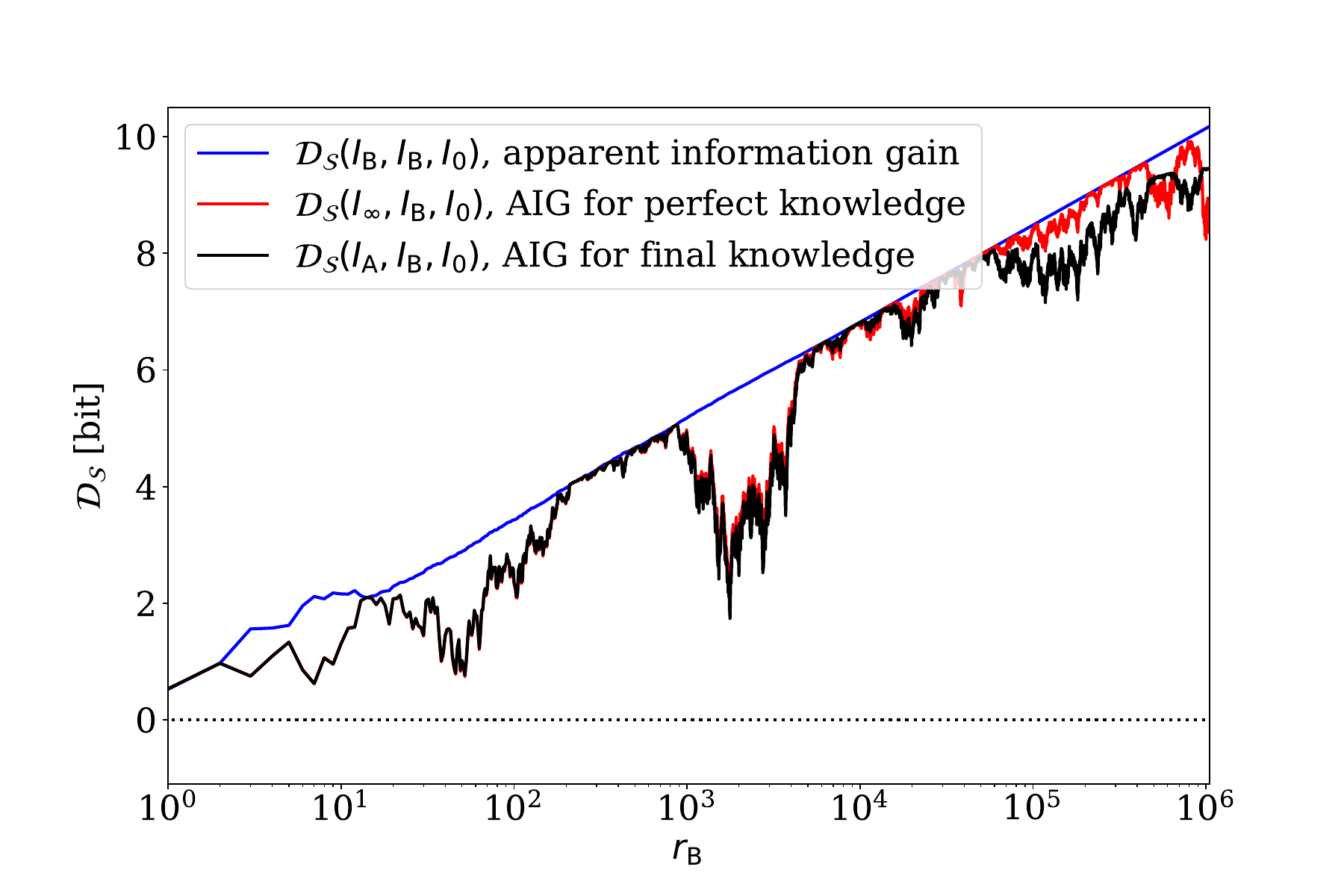}\includegraphics[viewport=40bp 20bp 800bp 510bp,clip,width=0.49\textwidth]{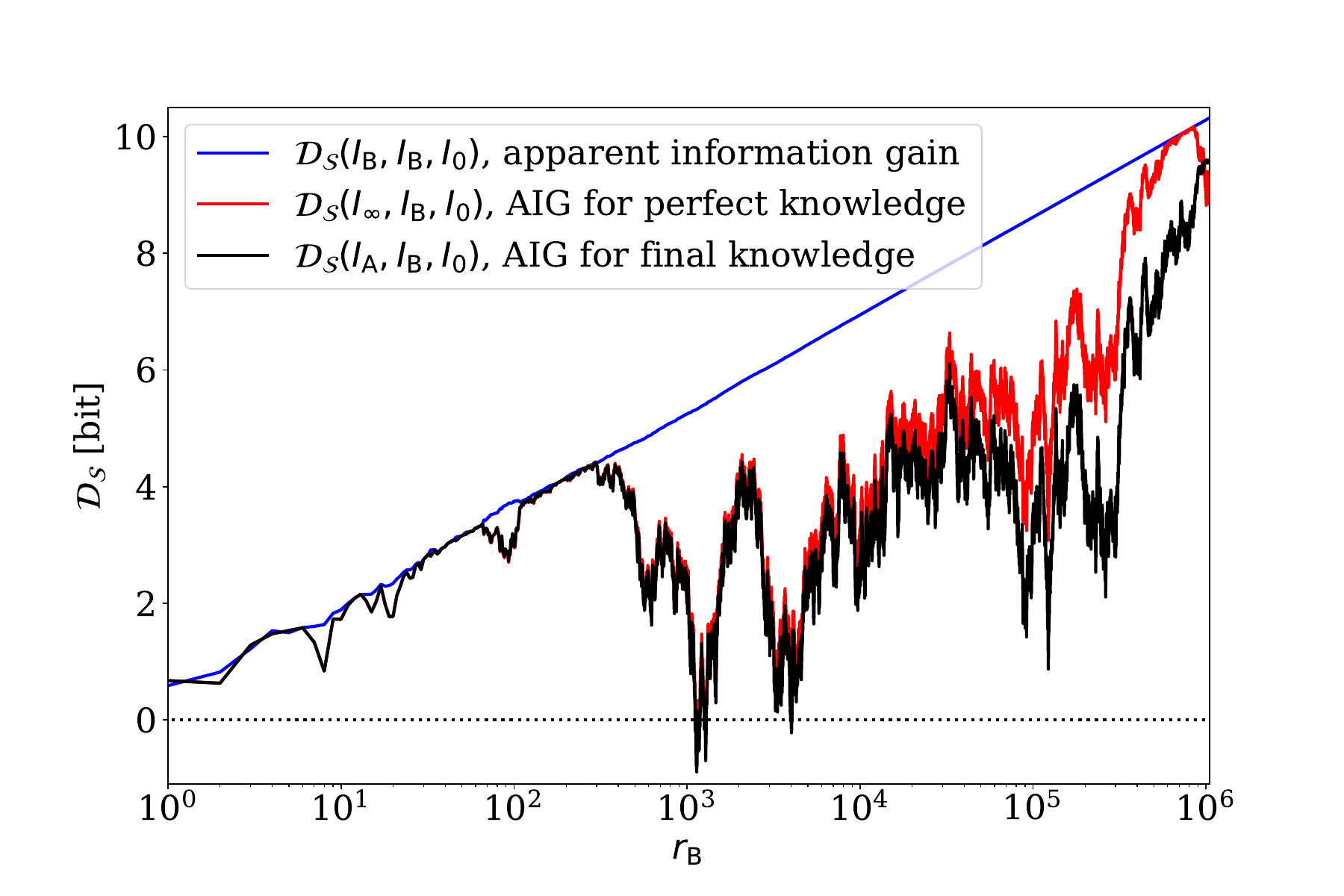}\caption{Various information gains in case $I_{\text{B}}$ represents incomplete
data usage, where the total dataset of $I_{\text{A}}$ has $r_{\text{A}}=2^{20}\approx10^{6}$
measurements of a scalar quantity, each with Gaussian noise that has
the variance of the prior uncertainty, $\sigma_{n}=\sigma_{s}$ (a
signal to noise ratio of one), for two different signal and data realizations
shown in the two panels. The black curve shows $\mathcal{D}_{\mathcal{S}}(I_{\text{A}},I_{\text{B}},I_{\text{0}})$,
the AIG with respect to the final state $I_{\text{A}}$ using all
data, the red line shows $\mathcal{D}_{\mathcal{S}}(I_{\text{\ensuremath{\infty}}},I_{\text{B}},I_{\text{0}})$,
the AIG with respect to a perfect knowledge of the ground truth, $I_{\infty}=(m_{\infty},\sigma_{\infty})=(s,0)$
or equivalently $\mathcal{P}(s|I_{\infty})=\delta(s-m_{\infty})$,
and the blue curve shows the apparent information gain $\mathcal{D}_{\mathcal{S}}(I_{\text{B}},I_{\text{B}},I_{\text{0}})=\mathcal{D}_{\mathcal{S}}(I_{\text{B}},I_{\text{0}})$
. \protect\label{fig:incompl_data}}
\end{figure*}
A special case of a imperfect Gaussian updates are those in which
only a part of the data available is used. Let 
\begin{equation}
d_{i}=s+n_{i}
\end{equation}
be the individual measurement equations of a repeated measurement
of a real quantity $s\in\mathbb{R}$ with prior $\mathcal{P}(s)=\mathcal{G}(s,S)$
with known uncertainty variance $S=\sigma_{s}^{2}$. We will investigate
how the AIG of the posterior changes if only the first $r_{\text{B}}$
of the $r_{\text{A}}$ measurements are used. The achieved gain will
in general depend on the data realization $d_{\text{A}}=(d_{1},\ldots\,d_{r_{\text{B}}},\ldots\,d_{r_{\text{\text{A}}}})^{\text{t}}=:(d_{\text{B}}^{\text{t}},d_{\text{A}\backslash\boldsymbol{\text{B}}}^{\text{t}})^{\text{t}}$.
The updated information states are $I_{\text{B}}=(d_{\text{B}}^{\text{t}},I_{0}^{\text{t}})^{\text{t}}$
and $I_{\text{A}}=(d_{\text{A}}^{\text{t}},I_{0}^{\text{t}})^{\text{t}}$,
the augmentations of $I_{\text{0}}$ with the initial and final data
vector, respectively.

The measurement equations can be brought into a vector notation, 
\begin{equation}
d_{\text{X}}=R_{\text{X}}s+n_{X},
\end{equation}
with $\text{X}\in\{\text{A,}\text{B}\}$. The responses $R_{\text{X}}=(1)_{i=1}^{r_{\text{X}}}$
are one-column matrices that turn the signal into the shape of a data
vector of length $r_{\text{X}}$. We assume the measurement noise
$n_{\text{X}}$ to be independent of the signal as well as identical
and independently distributed (iid) according to a zero centered Gaussian
with known variance $\sigma_{n}^{2}$ for each individual measurement.
Thus, the joint data and signal probability is
\begin{eqnarray}
\mathcal{\mathcal{P}}(d_{\text{X}},s|I_{\text{0}}) & = & \mathcal{G}(s,\sigma_{s}^{2})\mathcal{\,G}(d_{\text{X}}-R_{\text{X}}s,N_{X}),
\end{eqnarray}
where $N_{X}=\mathbb{1}_{r_{\text{X}}}\sigma_{n}^{2}$ is the noise
covariance with $\mathbb{1}_{r_{\text{X}}}$ denoting the $r_{\text{\text{X}}}\times r_{\text{\text{X}}}$
unit matrix. The posterior is then the so called Wiener filter posterior
distribution \cite{2009PhRvD..80j5005E}, 
\begin{eqnarray}
\mathcal{P}(s|d_{\text{X}},I_{\text{0}}) & = & \mathcal{G}(s-m_{\text{X}},D_{\text{X}})\text{, with}\\
D_{\text{X}} & = & \left(S^{-1}+R_{\text{X}}^{\text{t}}N_{\text{X}}^{-1}R_{\text{X}}\right)^{-1}\text{ }\nonumber \\
 & = & \left(\sigma_{s}^{-2}+r_{\text{X}}\sigma_{n}^{-2}\right)^{-1}\nonumber \\
 & = & \left[1+q\,r_{\text{X}}\right]^{-1}\sigma_{s}^{2}\\
m_{\text{X}} & = & D_{\text{X}}R_{\text{X}}^{\text{t}}N_{\text{X}}^{-1}d_{\text{X}}=\frac{\sigma_{n}^{-2}\sum_{i=1}^{r_{\text{X}}}d_{i}}{\sigma_{s}^{-2}+r_{\text{X}}\sigma_{n}^{-2}}\nonumber \\
 & = & \left[1+(q\,r_{\text{X}})^{-1}\right]^{-1}\overline{d_{\text{X}}}.
\end{eqnarray}
Here, $\overline{d_{\text{X}}}:=\frac{1}{r_{\text{X}}}\sum_{i=1}^{r_{\text{X}}}d_{i}$
is the data mean up to measurement $r_{\text{X}}$ and $q:=\sigma_{s}^{2}/\sigma_{n}^{2}$
is the signal-to-noise variance ratio of an individual measurements.
We observe that with increasing accumulated signal to noise, as expressed
by $q\,r_{\text{X}}=r_{\text{X}}\,\sigma_{s}^{2}/\sigma_{n}^{2}$,
the posterior mean gets closer to the data mean and that the remaining
uncertainty decreases.

We identify the signal prior with $\mathcal{P}(s|I_{\text{0}})\equiv\mathcal{G}(s,\sigma_{s}^{2})$,
so that $m_{\text{0}}=0$ and $D_{\text{0}}=\sigma_{s}^{2}$ and the
posteriors for $\text{X}\in\{\text{A,}\text{B}\}$ with 
\begin{eqnarray}
\mathcal{P}(s|I_{\text{X}}) & = & \mathcal{G}(s-m_{\text{X}},D_{\text{X}}).
\end{eqnarray}

The AIG is according to Eq.\ \ref{eq:GaussianRIG}
\begin{eqnarray}
\mathcal{D_{\mathcal{S}}}(I_{\text{A}},I_{\text{B}},I_{\text{0}}) & = & \mathinner{\color{violet}\frac{1}{2}}{\color{violet}\ln}\mathinner{\color{violet}\frac{|D_{\text{0}}|}{|D_{\text{B}}|}}+\frac{1}{2}\text{Tr}\left[\left({\color{blue}D_{\text{0}}^{-1}}\mathbin{\color{red}-}{\color{red}D_{\text{B}}^{-1}}\right)D_{\text{A}}\right]\nonumber \\
 &  & +\frac{1}{2}\left({\color{olive}\Delta_{\text{0}}^{\text{t}}D_{\text{0}}^{-1}\Delta_{\text{0}}}\mathbin{\color{orange}-}{\color{orange}\Delta_{\text{B}}^{\text{t}}D_{\text{B}}^{-1}\Delta_{\text{B}}}\right)\nonumber \\
 & = & \mathinner{\color{violet}\frac{1}{2}}{\color{violet}\ln}\mathinner{\color{violet}\frac{\sigma_{s}^{2}}{\left[1+q\,r_{\text{B}}\right]^{-1}\sigma_{s}^{2}}}\nonumber \\
 &  & +\frac{1}{2}\,\frac{{\color{blue}\sigma_{s}^{-2}}\mathbin{\color{red}-}{\color{red}\left[1+q\,r_{\text{B}}\right]\sigma_{s}^{-2}}}{1+q\,r_{\text{A}}}\sigma_{s}^{2}+\mathinner{\color{olive}\frac{1}{2}}{\color{olive}\sigma_{s}^{-2}m_{\text{A}}^{2}}\nonumber \\
 &  & \mathbin{\color{orange}-}{\color{orange}\frac{1}{2}\left(1+q\,r_{\text{B}}\right)\sigma_{s}^{-2}}\mathinner{\color{orange}\left(m_{\text{A}}-m_{B}\right)^{2}}\nonumber \\
 & = & \mathinner{\color{violet}\frac{1}{2}}{\color{violet}\ln}\mathinner{\color{violet}\left(1+q\,r_{\text{B}}\right)}-\frac{{\color{red}q\,r_{\text{B}}}}{2\,\left(1+q\,r_{\text{A}}\right)}\nonumber \\
 &  & +\frac{{\color{olive}m_{\text{A}}^{2}}\mathbin{\color{orange}-}{\color{orange}\left(1+q\,r_{\text{B}}\right)}\mathinner{\color{orange}\left(m_{\text{A}}-m_{B}\right)^{2}}}{2\sigma_{s}^{2}}.\label{eq:RIGincompleteMeasuremnt}
\end{eqnarray}
Here, we used that $\Delta_{0}:=m_{\text{A}}-m_{\text{0}}=m_{\text{A}}.$
The first two terms are independent of the data realization, but the
others are not.

Fig.\ \ref{fig:incompl_data} shows the evolution of the AIG, the
AIG for a perfect final knowledge, and the apparent information gain
for two data realizations. A number of observations can be made:
\begin{enumerate}
\item The overall trend is a logarithmic growth of all shown information
gains with the number of used measurements.
\item Unlucky data realizations can lead to temporary losses of already
gained information, even to a negative AIG.
\item The apparent information gain is always larger than the achieved ones.
It does not follow the decreases the other have under unlucky data
realizations. Its usage instead of the AIG therefore leads to an overestimation
of the information gain.
\end{enumerate}
\textbf{}\textbf{\textcolor{red}{}}

\subsection{Intractable posterior}

In case the accurate target distribution is intractable, the integral
in Eq.\ \ref{eq:RIG} to calculate the AIG can in general not be
performed analytically. Let us assume that $N$ posterior samples
$s_{i}\hookleftarrow\mathcal{P}(s|I_{\text{A}})$ with $i\in\{1,\ldots N\}$
are available, e.g. from a suitable Monte Carlo method \cites{1953JChPh..21.1087M}{1970Bimka..57...97H}{1987PhLB..195..216D}{2017arXiv170102434B},
and that prior $\mathcal{P}(s|I_{\text{0}})$ and approximate posterior
$\mathcal{P}(s|I_{\text{B}})$ are available analytically. Then the
AIG can be estimated by replacing the posterior average in Eq.\ \ref{eq:RIG}
by a sample average:
\begin{eqnarray}
\mathcal{D}_{\mathcal{S}}(I_{\text{A}},I_{\text{B}},I_{\text{0}}) & \approx & \frac{1}{N}\sum_{i=1}^{N}\ln\frac{\mathcal{P}(s_{i}|I_{\text{B}})}{\mathcal{P}(s_{i}|I_{\text{0}})}\nonumber \\
 & = & \left\langle \mathcal{H}(s_{i}|I_{\text{0}})-\mathcal{H}(s_{i}|I_{\text{B}})\right\rangle _{i}
\end{eqnarray}

If both are actually Gaussian distributions, as specified by Eq.\ \ref{eq:Gaussian},
then this sample average becomes
\begin{eqnarray}
\mathcal{D}_{\mathcal{S}}(I_{\text{A}},I_{\text{B}},I_{\text{0}}) & \approx & \left\langle \ln\frac{\mathcal{G}(s_{i}-m_{\text{B}},D_{\text{B}})}{\mathcal{G}(s_{i}-m_{\text{0}},D_{\text{0}})}\right\rangle _{i}\nonumber \\
 & = & \frac{1}{2}\left[\ln\frac{|D_{\text{0}}|}{|D_{\text{B}}|}\right.\nonumber \\
 &  & +\left\langle (s_{i}-m_{\text{0}})^{\dagger}D_{\text{0}}^{-1}(s_{i}-m_{0})\right\rangle _{i}\nonumber \\
 &  & \left.-\left\langle (s_{i}-m_{\text{B}})^{\dagger}D_{\text{B}}^{-1}(s_{i}-m_{\text{B}})^{\dagger}\right\rangle _{i}\right].\nonumber \\
\end{eqnarray}
Often only approximate posterior samples are available, as their generation
might have used the variational inference approximation \cites{2019arXiv190111033K}{2021Entrp..23..853F}.
These can still be used to provide the AIG of even more approximate
methods w.r.t.\ them. An application of this is left to future work.

In case the ideal posterior distribution $\mathcal{P}(s|I_{\text{A}}(d)):=\mathcal{P}(s|d,I_{\text{0}})$
for data $d$ is not even accessible via samples, but the ground truth
$s_{\text{true}}$ is, e.g.\ from a computer simulations of signals
and data, the AIG for an approximate knowledge state $I_{\text{B}}(d)$
can be calculated with respect to the perfect knowledge state $\mathcal{P}(s|I_{s=s_{\text{true}}})=\delta(s-s_{\text{true}})$
according to Eq.\ \ref{eq:ground-truth}. This allows to characterize
any method to construct approximate posterior distributions $\mathcal{P}(s|I_{\text{B}}(d))\approx\mathcal{P}(s|d)$,
if pairs of prior and likelihood samples $s_{i}\hookleftarrow\mathcal{P}(s|I_{\text{0}})$
and $d_{i}\hookleftarrow\mathcal{P}(d|s_{i},I_{\text{A}})$ can be
obtained. The sample averaged AIG of the perfect knowledge states,
\begin{eqnarray}
\!\! & \!\!\!\!\!\!\!\! & \langle\mathcal{D}_{\mathcal{S}}(I_{s=s_{i}},I_{\text{B}}(d_{i}),I_{\text{0}})\rangle_{i}\nonumber \\
\!\! & \!\!\!\!\approx\!\!\!\! & \langle\mathcal{D}_{\mathcal{S}}(I_{s=s'},I_{\text{B}}(d'),I_{\text{0}})\rangle_{(d',s'|I_{\text{0}})}\nonumber \\
\!\! & \!\!\!\!=\!\!\!\! & \langle\langle\mathcal{D}_{\mathcal{S}}(I_{s=s'},I_{\text{B}}(d'),I_{\text{0}})\rangle_{(s'|d',I_{\text{0}})}\rangle_{(d'|I_{\text{0}})}\nonumber \\
\!\! & \!\!\!\!=\!\!\!\! & \left\langle \int_{\mathcal{S}}\!\!\text{d}s'\,\mathcal{P}(s'|d,I_{\text{0}})\int_{\mathcal{S}}\!\!\text{d}s\,\delta(s-s')\,\ln\frac{\mathcal{P}(s|I_{\text{B}}(d))}{\mathcal{P}(s|I_{\text{0}})}\right\rangle _{(d|I_{\text{0}})}\!\!\!\!\nonumber \\
\!\! & \!\!\!\!=\!\!\!\! & \left\langle \int_{\mathcal{S}}\!\!\text{d}s'\,\mathcal{P}(s'|d,I_{\text{0}})\ln\frac{\mathcal{P}(s'|I_{\text{B}}(d))}{\mathcal{P}(s'|I_{\text{0}})}\right\rangle _{(d|I_{\text{0}})}\nonumber \\
\!\! & \!\!\!\!=\!\!\!\! & \left\langle \mathcal{D}_{\mathcal{S}}(I_{\text{A}}(d),I_{\text{B}}(d),I_{\text{0}})\right\rangle _{(d|I_{\text{0}})},\label{eq:sampling-the-AIG}
\end{eqnarray}
turns out to be the AIG with respect to the correct posterior knowledge
state $I_{\text{A}}(d)$ averaged over data realizations $d\hookleftarrow\mathcal{P}(d|I_{\text{0}})$.
This allows to characterize any method to construct approximate posteriors
$I_{\text{B}}(d)$ in terms of expected AIG.

\subsection{Attention}

Not every bit has the same value for the aims of the cognitive system.
There are quantities of higher relevance and quantities of lesser
importance. There are situations that are more important to know about
than others. In order to have the possibility to include a notion
of relevance into cognition processes, the concept of attention and
attention entropy can be defined \cite{2024AnP...53600334E}. An attention
function is a probability function that is modified by a weight function
$w:\mathcal{S\mapsto}\mathbb{R}_{0}^{+}$ and then normalized,
\begin{equation}
\mathcal{A}^{(w)}(s|I):=\frac{w(s)\,\mathcal{P}(s|I)}{\int_{\mathcal{S}}\text{d}s\,w(s)\,\mathcal{P}(s|I)}.
\end{equation}
A relative attention entropy is a relative entropy using attention
functions
\begin{eqnarray}
\mathcal{D}_{\mathcal{S}}^{(w)}(I_{\text{A}},I_{\text{B}}) & := & \int_{\mathcal{S}}\text{d}s\,\mathcal{A}^{(w)}(s|I_{\text{A}})\,\ln\frac{\mathcal{A}^{(w)}(s|I_{\text{A}})}{\mathcal{A}^{(w)}(s|I_{\text{B}})}\nonumber \\
 & = & \frac{\int_{\mathcal{S}}\text{d}s\,w(s)\,\mathcal{P}(s|I_{\text{A}})\,\ln\frac{\mathcal{P}(s|I_{\text{A}})\,}{\mathcal{P}(s|I_{\text{B}})}}{\int_{\mathcal{S}}\text{d}s\,w(s)\,\mathcal{P}(s|I_{\text{A}})}\nonumber \\
 &  & -\ln\frac{\int_{\mathcal{S}}\text{d}s\,w(s)\,\mathcal{P}(s|I_{\text{A}})}{\int_{\mathcal{S}}\text{d}s\,w(s)\,\mathcal{P}(s|I_{\text{B}})}.
\end{eqnarray}
Minimizing this for example with respect to $I_{\text{B}}$ can be
used by Alice to construct a message for Bob that take relevance as
encoded in $w(s)$ into account.

In the same way, an achieved attention gain can be defined,
\begin{eqnarray}
\mathcal{D}_{\mathcal{S}}^{(w)}(I_{\text{A}},I_{\text{B}},I_{0}) & := & \int_{\mathcal{S}}\text{d}s\,\mathcal{A}^{(w)}(s|I_{\text{A}})\,\ln\frac{\mathcal{A}^{(w)}(s|I_{\text{B}})}{\mathcal{A}^{(w)}(s|I_{\text{0}})}\nonumber \\
 & = & \frac{\int_{\mathcal{S}}\text{d}s\,w(s)\,\mathcal{P}(s|I_{\text{A}})\,\ln\frac{\mathcal{P}(s|I_{\text{B}})\,}{\mathcal{P}(s|I_{\text{0}})}}{\int_{\mathcal{S}}\text{d}s\,w(s)\,\mathcal{P}(s|I_{\text{A}})}\nonumber \\
 &  & -\ln\frac{\int_{\mathcal{S}}\text{d}s\,w(s)\,\mathcal{P}(s|I_{\text{B}})}{\int_{\mathcal{S}}\text{d}s\,w(s)\,\mathcal{P}(s|I_{\text{0}})},
\end{eqnarray}
which is a measure of the amount of attention achieved in a cognition.
The axiomatic derivation of this should be analogous to the ones given
in Sect.\ \ref{sec:Axiomatic-derivation} and in \cite{2024AnP...53600334E}
and is omitted here for brevity.

An attention fidelity can then be defined analogously,
\begin{equation}
\varepsilon_{\mathcal{S}}^{(w)}(I_{\text{A}},I_{\text{B}},I_{0})=\frac{\mathcal{D}_{\mathcal{S}}^{(w)}(I_{\text{A}},I_{\text{B}},I_{0})}{\mathcal{D}_{\mathcal{S}}^{(w)}(I_{\text{A}},I_{\text{A}},I_{0})}.
\end{equation}

\section{Sustainable data analysis\protect\label{sec:Sustainable-data-analysis}}

\subsection{Cognitive Efficiency}

With increasing costs of scientific experiments, observatories, and
the necessary computational efforts for their data analysis, the question
arises, how to optimize the cognitive efficiency of science, the amount
of scientific information obtained per invested money and other resources
\parencite{2024EPJST.tmp..399B}. We define cognitive efficiency as
\[
\text{CE}_{\text{\ensuremath{\mathcal{S}}}}(I_{\text{A}},I_{\text{B}},I_{0}):=\frac{\mathcal{D}_{\mathcal{S}}(I_{\text{A}},I_{\text{B}},I_{0})}{C(I_{\text{A}},I_{\text{B}},I_{0})},
\]
where $C(I_{\text{A}},I_{\text{B}},I_{0})$ denotes the costs associated
with performing the information update $I_{\text{0}}\rightarrow I_{\text{B}}$
in a situation where $I_{\text{0}}\rightarrow I_{\text{A}}$ would
be the optimal update. In general, methods with higher cognitive efficiency
should be preferred over such with lower efficiency.

For this argument to hold, both, numerator and denominator of cognitive
efficiency need to be carefully discussed. The achieved information
gain might not the only benefit of an update, as for example it can
have an educational, cultural, technological, or political dimension.
However, here we focus on the amount of AIG. When estimating the costs,
it often makes a significant difference whether only the computational
costs of an update are considered, or the costs to obtain the data
used in the update are also included into the cost budget. The former
is an appropriate approximation in case the data would be available
anyway and thus the data acquisition costs are vanishing small. The
latter must be used in case a dedicated investment was necessary to
obtain the data. From the perspective of the society, the data generation
costs should be included in any sustainability calculation.

As a consequence of this, the cognitive efficiency differs significantly
for different perspectives. To a scientist, who analyses a freely
available dataset, a less expensive method that has lower cognitive
fidelity can be more appealing, as it might maximize his cognitive
efficiency. From a societal perspective, however, the costs of producing
the data should be taken into account, rendering more accurate, but
usually computationally more expensive methods beneficial from a global
cognitive efficiency point of view, assuming of course that the higher
accuracy is beneficial.

In scientific practice, this gap in interests might often be closed
by the mechanisms of scientific publication. These usually require
that a reanalysis of data needs to have an increased cognitive fidelity
in comparison to earlier ones in order to be accepted by a peer reviewed
journal.

\subsection{Sustainable costs}

In order to decide from a sustainability perspective which of two
cognitive methods, say ``$\text{B}$'' and ``$\text{C}$'' should
be used to analyze data from a measurement device, we have to compare
their benefits and costs. As the benefit of an AIG does not need to
increase linearly with its size, the best way to compare two method
is not using the same dataset for both, but to request that each of
them is provided with a data set sized such that it leads to the same
(expected) AIG for each of them. This way, their benefits will be
identical, but their data acquisition and processing costs will differ,
which are usually easier to quantify.

Suppose for reaching a certain expected AIG level, a fraction of time
$f_{M}$ of a measurement facility is needed for method $M\in\{\text{B},\text{C}\}$.
The total cost of the facility be $C^{\text{facility}}$ and that
of the computational method be $C_{M}^{\text{comp}}$. Thus, the total
cost of this scientific result with this method is
\begin{equation}
C_{M}^{\text{total}}=f_{M}\,C^{\text{facility}}+C_{M}^{\text{comp}}.
\end{equation}
Method $\text{B}$ is then more economic than method $\text{C}$ if
$C_{\text{B}}^{\text{total}}<C_{\text{C}}^{\text{total}}$, as their
expected AIG and thus their societal benefits coincide. Thus, method
$\text{B }$should be preferred over $\text{C}$ if
\begin{eqnarray}
\Delta C^{\text{comp}} & < & \Delta f\,C^{\text{facility}}\text{, with}\nonumber \\
\Delta C^{\text{comp}} & := & C_{\text{B}}^{\text{comp}}-C_{\text{C}}^{\text{comp}}\text{ and}\nonumber \\
\Delta f & := & f_{\text{C}}-f_{\text{B}}.
\end{eqnarray}
As a consequence of this, computationally more accurate and thereby
more expensive data analysis methods can be more sustainable than
less accurate and thereby less expensive ones, in particular when
data acquisition costs are high. This argument is strengthened by
the observation we made in Sect.\ \ref{subsec:Incomplete-data-usage}
that the AIG tends to grow only as the logarithm of the size of a
data set. This means that a method $\text{C }$with a lower cognitive
fidelity might require a significant larger data set to reach the
same AIG. Obtaining this larger data set consumes a larger fraction
$f_{\text{C}}$ of the expensive facility time and thereby worsen
the sustainability of the computationally inexpensive method.

\subsection{Illustrative scenario}

To illustrate this, let us consider the fictitious scenario of a larger
research facility with a price tag of $C^{\text{facility}}=10^{9}\text{ Euro}$
for 10 years of operation. We imagine that the more expensive data
analysis method $\text{B}$ has a $20\,\%$ increased data fidelity
compared to method $\text{C}$, $\varepsilon_{\text{B}}(d_{\text{B}})=1.2\,\varepsilon_{\text{C}}(d_{\text{B}})$,
for a dataset $d_{\text{B}}$ that represents $|d_{\text{B}}|=10$
independent measurements of the quantity of interest. We assume that
this data set can be taken in a day, meaning $f_{\text{B}}=\nicefrac{\text{day}}{\text{decade}}\approx0.27\,\permil$.
We assume further, inspired by the observations in Sect.\ \ref{subsec:Incomplete-data-usage},
that AIG and therefore cognitive fidelity grow logarithmic with the
data size, $\varepsilon_{\text{B}}(d_{\text{B}})=a_{\text{B}}\ln(|d_{\text{B}}|)$
and $\varepsilon_{\text{C}}(d_{\text{C}})=a_{\text{C}}\ln(|d_{\text{C}}|)$
with $a_{\text{B}}=1.2\,a_{\text{C}}$. Thus, the requirement of matching
AIGs, $\varepsilon_{\text{B}}(d_{\text{B}})=a_{\text{B}}\ln(|d_{\text{B}}|)=a_{\text{C}}\ln(|d_{\text{C}}|)=\varepsilon_{\text{C}}(d_{\text{C}})$,
leads to $|d_{\text{C}}|=|d_{\text{B}}|^{\nicefrac{a_{\text{B}}}{a_{\text{C}}}}=|d_{\text{B}}|^{1.2}$.
This implies a by a factor $\nicefrac{f_{\text{C}}}{f_{\text{B}}}=\nicefrac{|d_{\text{C}}|}{|d_{\text{B}}|}=|d_{\text{B}}|^{\nicefrac{a_{\text{B}}}{a_{\text{C}}}-1}=10^{0.2}\approx1.5$
increased necessary measurement time for method $\text{C}$. The extra
computational costs of method $\text{B}$ would therefore amortize
if they are below
\begin{eqnarray}
\Delta C_{\text{max}}^{\text{comp}} & = & \left(f_{\text{C}}-f_{\text{B}}\right)\,C^{\text{facility}}\nonumber \\
 & = & \left(|d_{\text{B}}|^{\nicefrac{a_{\text{B}}}{a_{\text{C}}}-1}-1\right)f_{\text{B}}\,C^{\text{facility}}\\
 & \approx & 0.5\times0.27\,\permil\times10^{9}\text{ Euro}\nonumber \\
 & \approx & 135\,000\text{ Euro}.\nonumber 
\end{eqnarray}
This might even accommodate the personnel cost for the development
of method $\text{B}$, in particular, in case it can be used for more
than one of such measurements.

\section{Conclusion\protect\label{sec:Conclusions}}

\subsection{Summary}

The need to quantify imperfect cognitive operations, be them biological
or computational, be them communication, inference, or memorization,
led us to introduce and axiomatically derive the concept of achieved
information gain (AIG) as the optimal possible gain minus the remaining
gain after the imperfect cognition. We showed that AIG has many of
the properties required to characterize imperfect cognitive operations
and allows us to define cognitive fidelity and cognitive efficiency.
Furthermore, we discussed its relation to several other information
measures and showed how those can also be turned into achieved gains
as well by distinguishing the reached knowledge state of an update
from the ideally reached one.

We examined analytically illustrative scenarios with Gaussian and
various non-Gaussian probabilities and showed how to calculate the
AIG numerically in case of intractable distributions via sampling.
We showed that the practice of enlarging uncertainties in case of
an unaccounted error in the mean value of an update is encouraged,
as it can turn an otherwise negative AIG into a positive one. For
repeated measurements of a quantity, numerical experiments indicate
that on average the AIG grows with the logarithm of the data set size
obtained, interrupted by episodes of unfortunate data sequences with
significantly reduced AIG. The apparent information gain is insensitive
to such unlucky data realizations and therefore can be misleading.

Calculating an AIG in practice has the computational complexity of
computing the relative entropy (aka Kullback-Leibler divergence).
It has -- however -- the additional complication that an ideal update
probability $\mathcal{P}(s|I_{\text{A}})$ is required, in addition
to the initial and updated probabilities $\mathcal{P}(s|I_{\text{0}})$
and $\mathcal{P}(s|I_{\text{B}})$, respectively. For example in Bayesian
knowledge updates after receiving data $d\in D$, the posterior $\mathcal{P}(s|I_{\text{A}}(d)):=\mathcal{P}(s|d,I_{\text{0}})$
might be intractable. $\mathcal{P}(s|I_{\text{0}})$ would be the
prior and $\mathcal{P}(s|I_{\text{B}}(d))$ an approximate posterior
that should be characterized via AIG. But if prior and likelihood
allow for sample generation, $s_{i}\hookleftarrow\mathcal{P}(s_{i}|I_{\text{0}})$
and $d_{i}\hookleftarrow\mathcal{P}(d_{i}|s_{i},I_{\text{0}})$, an
expected AIG of the method $I_{\text{B}}:D\rightarrow\mathcal{I}$
can be estimated via a simple sample average,
\begin{equation}
\left\langle \mathcal{D}_{\mathcal{S}}(I_{\text{A}}(d),I_{\text{B}}(d),I_{\text{0}}\right\rangle _{(d|I_{\text{0}})}\approx\left\langle \ln\frac{\mathcal{P}(s|I_{\text{B}}(d_{i}))}{\mathcal{P}(s|I_{\text{0}})}\right\rangle _{i},
\end{equation}
as follows from Eqs.\ \ref{eq:ground-truth} and \ref{eq:sampling-the-AIG}.

Finally, we illustrated how AIG can be used to help to decide the
trade-off between accuracy and computational complexity for expensive
data of large research facilities. In deciding which computational
method is more sustainable, the concept of AIG allows us to identify
how much longer a facility needs to be used to generate as informative
data for a method with lower cognitive fidelity, and thus sets the
scale at which the high fidelity method pays off despite being potentially
more computational expensive.

\subsection{Outlook}

We hope that the concept of AIG will find ample applications in quantifying
and understanding technical, biological, psychological and sociological
information processing. It should help the design of cognitive efficient
data processing systems, for example for the energetically expensive
processing of the ever increasing data streams generated by the growing
set of scientific, industrial, and private sensors, detectors, and
telescopes \parencite{2024EPJST.tmp..399B}. It can help to understand
the trade-offs that shaped the evolution of existing biological, psychological,
or sociological information processing systems. And as an information
``distance'' measure that depends on three locations, the initial,
the reached, and the ideal one, AIG provides insight into the geometrical
properties of approximate cognitive operations, which might become
relevant for example in understanding the operations of artificial
intelligence systems.

\subsection*{Acknowledgments}

This research was inspired by the Workshop ``Sustainability in the
Digital Transformation of Basic Research on Universe and Matter''
in 2023, which was supported by the Ministry of Innovation, Science,
and Research of the State of North Rhine-Westphalia, and by the Federal
Ministry of Education and Research (BMBF) in Germany through the ErUM-Data-Hub
project 05D21PA1. TE thanks the workshop organizers and participants
for useful discussion. The author acknowledged helpful comments on
the manuscript by Andreas Popp, Viktoria Kainz, Johannes Harth-Kitzerow,
Matteo Guardiani, Fabrizia Guglielmetti, Robert Zimmermann, and an
anonymous referee.

\printbibliography

\end{document}